\documentclass[10pt,journal,compsoc]{IEEEtran}
\usepackage{amsmath}
\usepackage{amssymb}
\usepackage{algorithmic}

\usepackage{array}
\newcolumntype{P}[1]{>{\centering\arraybackslash}p{#1}}
\usepackage{algorithm2e}
\usepackage{enumitem}
\usepackage{booktabs,caption}
\usepackage[flushleft]{threeparttable}

\usepackage{booktabs}
\usepackage{tabularx}
\usepackage{subfig}
\usepackage{color, colortbl}
\definecolor{Gray}{gray}{0.9}

\usepackage{stfloats}
\usepackage{float}

\usepackage{graphicx}
\usepackage[table]{xcolor}
\usepackage{subfig}
\usepackage{url}
\usepackage{breakurl}
\usepackage[breaklinks]{hyperref}
\usepackage{hyperref}
\usepackage{multirow}
\usepackage{color}

\usepackage[font=small,labelfont=bf]{caption}
\newcommand{\kreminder}[1]{\textcolor{red}{\textbf{kurt: #1}}}
\renewcommand{\kreminder}[1]{{}}

\newcommand{\modification}{\color{blue}}
\newcommand{\modificationend}{\color{black}}
\newcommand{\modif}{\color{red}}
\newcommand{\modifend}{\color{black}}

\usepackage{epstopdf}

\usepackage{cite}
\ifCLASSINFOpdf
\else
\fi

\usepackage{xcolor}
\usepackage{fancyhdr}
\begin{document}

\title{A Context-aware Framework for Detecting Sensor-based Threats on Smart Devices}

\author{Amit~Kumar~Sikder,
        Hidayet~Aksu,
        and~A.~Selcuk~Uluagac
\IEEEcompsocitemizethanks{\IEEEcompsocthanksitem A. K. Sikder, H. Aksu, and A. S. Uluagac are with the Cyber-Physical Systems Security Lab (CSL), Electrical and Computer Engineering Department, Florida International University, Miami, Florida, USA. \protect\\
E-mail: {asikd003, haksu, suluagac}@fiu.edu
}}

\IEEEtitleabstractindextext{%
\begin{abstract}
Sensors (e.g., light, gyroscope, accelerometer) and sensing-enabled applications on a smart device make the applications more user-friendly and efficient. However, the current permission-based sensor management systems of smart devices only focus on certain sensors and any App can get access to other sensors by just accessing the generic sensor Application Programming Interface (API). In this way, attackers can exploit these sensors in numerous ways: they can extract or leak users' sensitive information, transfer malware, or record or steal sensitive information from other nearby devices. In this paper, we propose 6thSense, a context-aware intrusion detection system which enhances the security of smart devices by observing changes in sensor data for different tasks of users and creating a contextual model to distinguish benign and malicious behavior of sensors. 6thSense utilizes three different Machine Learning-based detection mechanisms (i.e., Markov Chain, Naive Bayes, and LMT). We implemented 6thSense on several sensor-rich Android-based smart devices (i.e., smart watch and smartphone) and collected data from typical daily activities of 100 real users. Furthermore, we evaluated the performance of 6thSense against three sensor-based threats: (1) a malicious App that can be triggered via a sensor, (2) a malicious App that can leak information via a sensor, and (3) a malicious App that can steal data using sensors. Our extensive evaluations show that the 6thSense framework is an effective and practical approach to defeat growing sensor-based threats with an accuracy above 96\% without compromising the normal functionality of the device. Moreover, our framework reveals minimal overhead. 
\end{abstract}
}

\maketitle

\IEEEdisplaynontitleabstractindextext

\IEEEpeerreviewmaketitle

\thispagestyle{fancy}

\IEEEraisesectionheading{\section{Introduction}\label{sec:introduction}}

\IEEEPARstart{S}{mart} devices such as smartphones and smartwatches 
have become omnipresent in every aspect of human life. Nowadays, the role of smart devices is not limited to making phone calls and messaging only. They are integrated into
various applications from home security to health care to
military to smart city~\cite{chan2009smart, babun2018iotdots, sikder2018iot}. Since smart devices seamlessly integrate the physical world with the cyber world via their sensors (e.g., light, accelerometer, gyroscope, etc.), they provide more efficient and user-friendly applications~\cite{s131217292, 217632, sikder2018iot}. In a way, sensors are eyes, ears, skin of the device to the physical world as five sensing organs are to the human beings.

While the number of applications using different sensors is increasing and new devices offer more sensors, the presence of sensors have opened novel ways to exploit the smart devices~\cite{6997498}. Attackers can exploit the sensors in multiple ways~\cite{6997498}: They can trigger an existing malware on a device with a simple flashlight~\cite{Hasan:2013:SCH:2484313.2484373}; they can use a sensor (e.g., light sensor) to leak sensitive information; using motion sensors, attackers can record or steal sensitive information from other nearby devices (e.g., computers, keyboards) or people~\cite{Halevi:2012:CLK:2414456.2414509, maiti2016smartwatch}. They can even transfer a specific malware using sensors as a communication channel~\cite{6997498}. Such \textit{sensor-based} threats become more serious with the rapid growth of Apps utilizing many sensors~\cite{ICS-CERT}.

In fact, these sensor-based threats highlight the flaws of existing sensor management systems used by smart devices. Specifically, Android sensor management system relies on permission-based access control, which considers only a few sensors (i.e., microphone, camera, and GPS)\footnote{IOS, Windows, and Blackberry also have similar permission-based sensor management systems. In this work, we focus on Android due to its open-source nature and high market share~\cite{androidmarket}}. Android asks for access permission (i.e., with a list of permissions) only while an App is being installed for the first time. Once this permission is granted, the user has no control over how the listed sensors and other sensors (not listed) will be used by the specific App. Moreover, using non-listed sensors is not considered as a violation of security and privacy in Android. For instance, any App is permitted to access to motion sensors by just accessing the \textit{sensor} API. Access to motion sensors is not controlled in Android.

Existing studies have proposed enhanced access control mechanisms for some of the sensors, but these enhancements do not cover all the sensors of a smart device~\cite{203854, Xu:2015:SPS:2699026.2699114}. Some proposed solutions introduced trusted paths on top of the existing security mechanisms for controlling information flow between sensors and Apps, but these are also App-specific solutions and depend upon explicit user consent. Thus, introducing additional permission controls for sensors of a smart device will not mitigate the risk of all sensor-based threats as they are App-specific and address only data leakage risks. Some attacks may not abuse sensors directly; instead, they may use sensors as side-channels to activate another malware~\cite{6997498}. Albeit useful, existing security schemes overlook these critical threats which directly impact the security and privacy of the smart device ecosystem. Moreover, although sensors on smart devices seem to work independently from each other, a task or activity on a smart device may activate more than one sensor to accomplish the task. Hence, it is necessary to secure all the different sensors on a smart device and consider the \textit{context} of the sensors in building any solution against the sensor-based threats.

\par In order to address the sensor-based threats, in this paper, we present a novel intrusion detection (IDS) framework called \textit{6thSense}, as a comprehensive security solution for sensor-based threats for smart devices. The proposed framework is a \textit{context-aware IDS} and is built upon the observation that for any user activity or task (e.g., texting, making calls, browsing, driving, etc.), a different, but a specific set of sensors becomes active. In a context-aware setting, the 6thSense framework is aware of the sensors activated by each activity or task. 6thSense observes sensors data in real time and determines the current use context of the device according to whether the current sensor use is malicious or not. 6thSense is context-aware and correlates the sensor data for different user activities (e.g., texting, making calls, browsing, etc.) on the smart devices and learns how sensors' data correlates with different activities. As a detection mechanism, 6thSense observes sensors' data and checks against the learned behavior of the sensors. In 6thSense, the framework utilizes several Machine Learning-based detection mechanisms to catch sensor-based threats including Markov Chain, Naive Bayes, and a set of other ML algorithms (e.g., PART, Logistic Function, J48, LMT, Hoeffding Tree, and Multilayer Perception). In this paper, we present the design of 6thSense on different Android devices (smartphone and smart watch) because of its open-source nature, large market share~\cite{androidmarket}, and rich set of sensors. To evaluate the efficiency of the framework, we tested it with data collected from real users (100 different users, 16 different typical daily activities for smartphone and smart watch~\cite{activity} including 153600 and 307200 different event-state information, respectively). We also evaluated the performance of 6thSense against three different sensor-based threats and finally analyzed its overhead. Our evaluation shows that 6thSense can detect sensor-based attacks with an accuracy and F-Score over 96\%. Also, our evaluation shows a minimal overhead on the utilization of the system resources. Note that, this work is an extension of our previous work~\cite{203854}. We significantly improved the framework from our prior work and implemented 6thSense on smart watch and smart phone. We also evaluated the performance with new user data and analyzed the performance overhead in further detail. 

\textit{\textbf{Contributions:}} In summary, the main contributions of this paper are threefold---
\begin{itemize}
\item \textit{First}, the design of 6thSense, a context-aware IDS to detect sensor-based threats utilizing multiple machine learning based models from Markov Chain to  Naive Bayes to LMT.   

\item \textit{Second}, the extensive performance evaluation of 6thSense with real user experiments over 100 users for different smart devices (smartphone and smart watch). 

\item \textit{Third}, testing of 6thSense against three different sensor-based threats.   

\end{itemize}

\textit{\textbf{Organization:}} The rest of the paper is organized as follows: we give an overview of sensor-based threats and existing solutions in Section 2. In section 3, we briefly discuss the \textit{Android's} sensor management system. Adversary model, design features, and assumptions for 6thSense are briefly discussed in Section 4. Different detection techniques used in our framework are described in Section 5. In Section 6, we provide a detailed overview of 6thSense including its different components and in Section 7, we evaluate its effectiveness by analyzing different performance metrics. Finally, we conclude this paper in Section 8.

\vspace{-0.5cm}
\section{Related Work}\label{sec:RelatedWork}
\vspace{-2pt}
\textit{Sensor-based threats}~\cite{6997498} on smart devices have become more prevalent than before with the use of different sensors such as user's location, keystroke information, etc. Several works \cite{6654855, acar2018peek} have investigated the possibility of these threats and presented different potential threats in recent years. Some interesting sensor-based threats are explained below.

One of the most common threats is keystroke inference in smart devices. When a user types in the keyboard, motion sensor readings (i.e., accelerometer and gyroscope) change accordingly~\cite{cai2012practicality}. As different keystrokes yield different, but specific values in motion sensors, typing information on on-screen keyboard can be inferred from an unauthorized sensor such as motion sensor data or its patterns collected either in the device or from a nearby device can be used to extract users' input in smart devices~\cite{sikder2018survey, shen2015input, 7113464, Wang:2015:MML:2789168.2790121}. The motion sensor data can be analyzed using different  techniques (e.g., machine learning, frequency domain analysis, shared-memory access, etc.) to improve the accuracy of inference techniques such as~\cite{Aviv:2012:PAS:2420950.2420957, Miluzzo:2012:TYF:2307636.2307666, Ping:2015:TIL:2766498.2766511,7605458, maiti2016smartwatch}. Another form of keystroke inference threat can be performed by observing only gyroscope data. Smart devices have a feature of creating vibrations while a user types on the touchpad. The gyroscope is sensitive to this vibrational force and it can be used to distinguish different inputs given by the users on the touchpad \cite{Narain:2014:SLK:2627393.2627417,Cai:2011:TIK:2028040.2028049, 184479}. Recently, ICS-CERT also issued an alert for accelerometer-based attacks that can deactivate any device by matching vibration frequency of the accelerometer \cite{ICS-CERT, son2015rocking}. 

Light sensor readings also change while a user types on smart devices;  hence, the user input in a smart device can be inferred by differentiating the light sensor data in normal and typing modes~\cite{Spreitzer:2014:PSE:2666620.2666622}. The light sensor can also be used as a medium to transfer malicious code and trigger message to activate a malware~\cite{Hasan:2013:SCH:2484313.2484373}. The audio sensor of a smart device can also be exploited to launch different malicious attacks (e.g., information leakage, eavesdropping, etc.) on the device. Attackers can infer keystrokes by recording tap noises on touchpad \cite{FooKune:2010:TAP:1866307.1866395}, record conversation of users \cite{schlegel2011soundcomber}, transfer malicious code to the device \cite{6654855}, or even replicate voice commands used in voice-enabled different Apps like \textit{Siri}, \textit{Google Voice Search}, \textit{Amazon echo}, \textit{Google Smart Home} etc.~\cite{Diao:2014:YVA:2666620.2666623}. Cameras of different smart devices can also be used to covertly capture screenshot or video and to infer information about surroundings or user activities \cite{Simon:2013:PSI:2516760.2516770, Meng:2015:CMI:2732198.2732205, Shukla:2014:BYH:2660267.2660360}. GPS of a smart device can be exploited to perform a false data injection attack on smart devices and infer the location of a specific device. 

\par \textbf{\textit{Solutions for sensor-based threats:}} Although researchers identified different sensor-based threats in recent years, no complete security mechanism has been proposed that can secure sensors of a smart device. Most of the proposed security mechanisms for smart devices are related to anomaly detection at the application level which are not built with any protection against sensor-based threats~\cite{Wu:2014:DDA:2663761.2664223}. On the other hand, different methods of intrusion detection have been proposed for wireless sensor networks (WSN)~\cite{pongaliur2008securing}, but they are not compatible with smart devices.  
Xu et al. proposed a privacy-aware sensor management framework for smartphones named \textit{Semadroid}~\cite{Xu:2015:SPS:2699026.2699114}, an extension to the existing Android sensor management system where users could monitor sensor usage of different Apps and invoke different policies to control sensor access by active Apps on a smartphone. Maiti et al. proposed a real-time activity detection framework to identify user activity on a smart device using motion sensor and allow motion sensor access based on the detected activity~\cite{maiti2016smartwatch}. Petracca et al. introduced \textit{AuDroid}, a SELinux-based policy framework for Android smartphones by performing behavior analysis of microphones and speakers~\cite{Petracca:2015:APA:2818000.2818005}. AuDroid controls the flow of information in the audio channel and notifies users whenever an audio channel is requested for access. An extension of this work is \textit{AWARE}, an authorization framework to secure privacy-sensitive sensors from malicious applications \cite{aware}. \textit{AWARE} considers both application requests and user interface to identify malicious user inputs in operation bindings for microphone and camera. Jana et al. proposed \textit{DARKLY}, a trust management framework for smartphones which audits applications of different trust levels with different sensor access permissions \cite{jana2013scanner}. Darkly  scans for vulnerability in the source code of an application and tries to modify the run-time environment of the device to ensure the privacy of sensor data. \par

\textbf{\textit{Differences from the existing solutions:}} 
Though there is no direct comparable work to compare 6thSense with, differences between existing solutions and our framework can be noted as follows: The main limitation of Semadroid~\cite{Xu:2015:SPS:2699026.2699114} is that the proposed solution is only tested against a similar type of attack scenario (information leakage by a background application). Semadroid also does not provide any extensive performance evaluation for the proposed scheme. Finally, this work depends on user permissions to fully enforce an updated policy on the sensor usage which is vulnerable as users might unknowingly approve the sensor permissions for malicious Apps. Real-time activity detection proposed by Maiti et al. considers motion sensors to identify user activity on a smart device which is only effective against keystroke inference~\cite{maiti2016smartwatch}. In Darkly \cite{jana2013scanner}, the proposed framework is not tested against any sensor-based threats. Audroid presented a policy enforced framework to secure only the audio channels of a smart device. Albeit useful, similar to the others, this work does not consider other sensor-based threats, either. More recent work \textit{AWARE} also considers selective sensors (e.g., camera and microphone) to identify malicious sensor accesses of the applications \cite{aware}. \textit{Compared to these prior works, 6thSense provides a comprehensive coverage to all the sensors in a smart device and ensures security against different types of sensor-based threats with high accuracy.}

\vspace{-10pt}
\section{Sensor Management in Smart Devices}
\begin{figure}[tb]
  \centering
  \framebox{  
    \includegraphics[height=6cm, width=5.5cm]{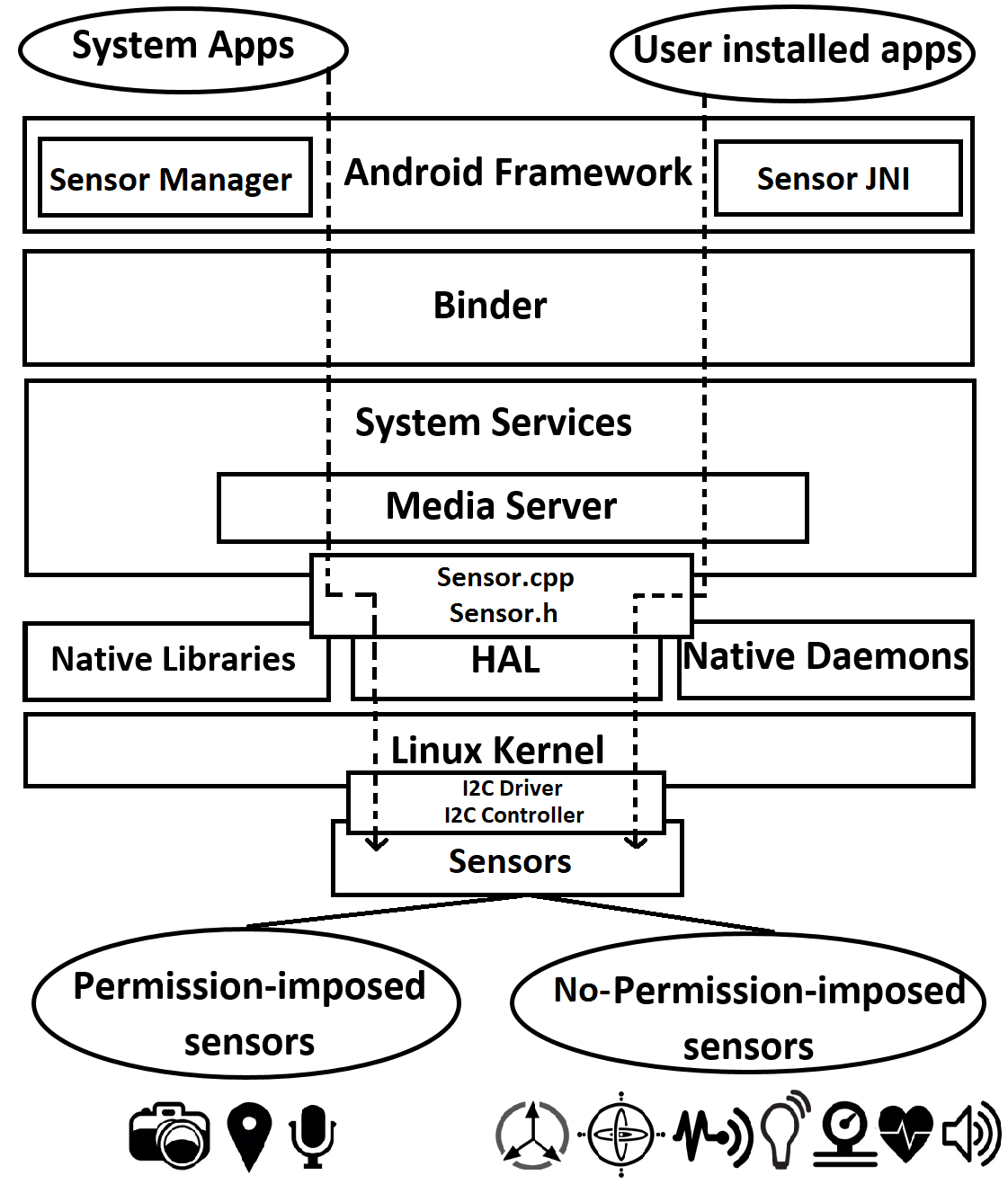}
    }
      \caption{Android Sensor Management Architecture.}
      \label{overview}
      \vspace{-0.5cm}
\end{figure}
Modern smart devices perform app-based operations which create a many-to-many relationship between sensors and Apps. Smart devices use more than one sensor to perform a task and hence, it is impractical to install an independent management system for each sensor. Smart device operating systems (OS) address this requirement by implementing centralized sensor management systems to manage and ensure secure data acquisition from all the sensors. In this section, we discuss sensor management systems of smart device OSes and articulate important deficiencies of the existed sensor management systems.

\noindent\textit{\textbf{Sensors in Smart Devices:}}
Most of the current smart device OSes offer permission-based sensor management system to control sensor access and data flow. According to the associated permissions imposed by the OSes, sensors in smart devices can be categorized in two groups - permission-imposed and no permission-imposed sensors.
\begin{itemize}
    \item \textit{Permission-imposed Sensor (PS):} Permission-imposed sensors are those which need explicit user permission to be accessed by an App. In smart devices, GPS, camera, and microphone are considered as permission-imposed sensor.
    \item \textit{No Permission-imposed sensors (NPS):} No-permission-imposed sensors can be defined as sensors that do not need any user permission explicitly to be  accessed by an App. Smart devices can have a wide range of no permission-imposed sensors such as accelerometer, gyroscope, proximity sensor, light sensor, etc.
\end{itemize}

\textit{\textbf{Sensor Management Systems in Smart Devices:}}
To understand sensor management systems in smart devices, we briefly explain sensor management in Android OS. In Figure~\ref{overview}, we present how Android handles access to different sensors by Apps (installed by the user) and system Apps (installed automatically by Android). Whenever an App wants to access sensors, it sends a request to \textit{SensorManager} via Software Development Kit (SDK) API which then registers the App to a corresponding sensor~\cite{milette2012professional}. If more than one App tries to access the same sensor, the SDK API runs a multiplexing process which enables different Apps to be registered in the same sensor. Hardware Abstraction Layer (HAL) works as an interface to bind the sensor hardware with the device drivers in Android. HAL has two parts: \textit{Sensors.h} works as HAL interface and \textit{Sensors.cpp} works as the HAL implementation. Through the HAL library, different applications can communicate with the underlying Linux kernel to read and write files associated with sensors. Also, the user permission for sensor access permission is declared inside the \textit{AndroidManifest.xml} file of an App and once the user accepts the permission, that App can have access to the corresponding permission-imposed sensors and other \textit{no-permission imposed sensors} even without any explicit approval from the users. 

\textit{\textbf{Limitations of Current Sensor Management Systems:}} Present versions of different smart device OSes (i.e., Android, iOS, Windows OS, Blackberry OS, etc.) do not comprise of any security mechanism to manage the information flow from sensors or among them. Most of the OSes offer a permission-based sensor management system to control sensor access and data flow between the application and sensor. OSes ask for user-permission for only selective set of sensors (e.g., GPS, camera, and microphone) at the time of installation or first use of an App. Users may grant access to sensitive sensors implicitly without knowing the actual intents of the App. Moreover, an App can get access to any no permission-imposed sensors by just accessing the \textit{sensor} API. Also, one task may need more than one sensor, but protecting only one sensor is not a viable design. The lack of ability to secure the information flow between the sensors and Apps and a holistic view into the utilization of sensors can lead to different malicious scenarios like information leakage, eavesdropping, etc.

\vspace{-10pt}
\section{Adversary Model and Assumptions}\label{sec:Adversarymodel}

In this section, we discuss different threats that may abuse sensors to execute malicious activities on a smart device. Different design features and assumptions are also explained in this section.

\textit{\textbf{Adversary Model:}}
For this work, we consider the following sensor-based threats similar to \cite{6997498}: 
\begin{itemize}
\item \textit{Threat 1-Triggering a malicious App via a sensor.} A malicious App can exist in the smart device which can be triggered by sending a specific sensory pattern or message via sensors. 
\item \textit{Threat 2-Information leakage via a sensor.} A malicious App can exist in the device which can leak information to any third party using sensors. 
\item \textit{Threat 3-Stealing information via a sensor.} A malicious App can exist in the device which can exploit the sensors of a smart device and start stealing information after inferring a specific device mode (e.g., sleeping). 
\end{itemize}

In this paper, we cover these three types of malicious sensor-based threats. To build our adversary model, we consider any component on a smart device that interacts with the physical world as a sensor \cite{Petracca:2015:APA:2818000.2818005}. We designed specific malware to represent above-mentioned threats and test our proposed framework against these malware.  

\noindent\textit{\textbf{Design Assumptions and Features:}}
In designing a comprehensive security scheme like 6thSense for sensor-based threats, we note the following design assumptions and features:
\begin{figure}[h!]
  \centering
    \includegraphics[width=9cm,height=4.5cm]{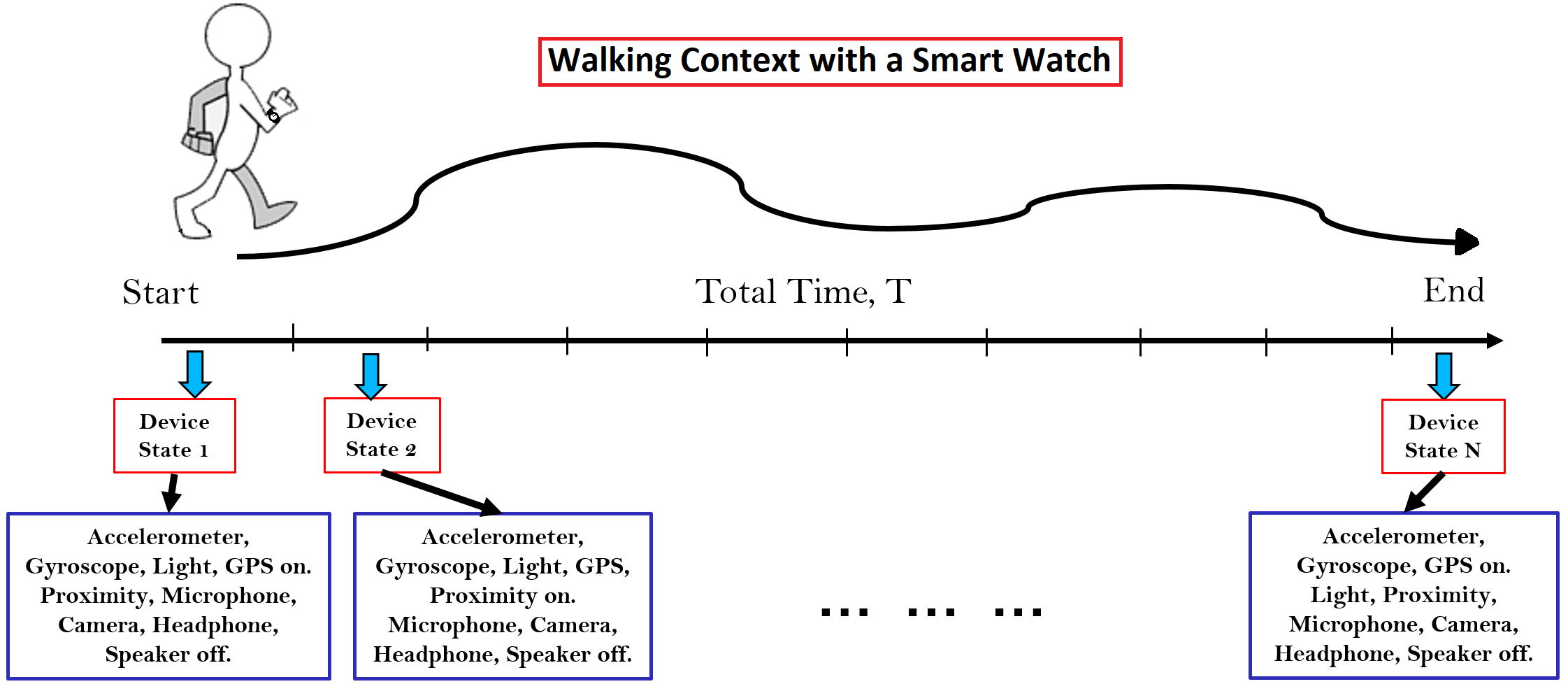}
    \vspace{-0.6cm}
    \caption{Context-aware model for 6thSense.}
    \vspace{-0.6cm}
      \label{context}
\end{figure}
\begin{itemize}
\item \textit{Context Awareness:} The main feature of 6thSense is context awareness which refers to the ability to sense the physical environment and adapt its operations accordingly in realistic cases~\cite{thyagaraju2012design}. 6thSense builds a context-aware model by observing the sensors' behaviors on a smart device in different usage scenarios. When a user is performing a task on a smart device, several sensors (i.e., accelerometer, gyroscope, light sensor, etc.) may remain active. This active state of different sensors is not constant and can change over time. This shifting in sensor's state over time should be considered correctly to understand the context of an activity. 6thSense divides the total execution time of an activity into smaller times and observes the sensors' states (on/off) over a short time span. Thus, whenever a sensor state is changed, 6thSense can understand the context and take a decision according to the context. For example, while a user is walking with a smartphone on his hand, several sensors (i.e., accelerometer, gyroscope, light sensor, etc.) remain active. If we divide the time of the activity in smaller times, we can see different sets of sensors active for different sensor states (Figure~\ref{context}). In this way, 6thSense considers all device states to understand the context of the activity and differentiate between benign and malicious activities. 
\item \textit{Sensor co-dependence:} A sensor in a smart device is normally considered as an independent entity on the device. Thus, one sensor does not know what is happening in another sensor. However, 
in this work, given an activity, we consider sensors as co-dependent entities on a device instead of independent entities. The reason for this stems from the fact that for each user activity or task on a smart device, a specific set of sensors remains active. For example, if a user is walking with a phone in hand, motion sensors (i.e., gyroscope, accelerometer), the light sensor, GPS will be active. On the contrary, if the user is walking with the phone in the pocket or bag,  instead of the light sensor, the proximity sensor will remain active. Thus, a co-dependent relationship exists  between sensors while performing a specific task. Each activity uses different, but specific set of sensors to perform the task efficiently. Hence, one can distinguish the user activity by observing the \textit{context of the sensors} for a specific task. 6thSense uses the context of all the sensors to distinguish between normal user activities and malicious activities. In summary, \textit{sensors in a smart device are individually independent, but per activity-wise dependent} and 6thSense considers the context of the activities in its design. 
\item \textit{Adaptive sensor sampling:} Different sensors have different sampling frequencies. To monitor all the sensor data for a specific time, a developed solution must consider and sample the sensor data correctly. 6thSense considers sampling the sensor data over a certain time period instead of individual sensor frequencies which mitigates any possible error in processing of data from different sensors. 6thSense collects each sensor data separately and samples the data according to their corresponding frequencies. These sample data are merged together to build contexts of different user activities in smart devices. 
\item \textit{Faster computation:} Modern high precision sensors on smart devices have high resolution and sampling rate. As a result, sensors provide large volume of data even for a small time interval. A solution for sensor-based threats should quickly process any large data from different sensors in real time while ensuring a high detection rate. To address this, we use different machine learning algorithms as detection techniques of 6thSense which are proven simple and fast techniques.
\item \textit{Real-time monitoring:} 6thSense provides real-time monitoring to all the sensors which mitigates the possibility of data tempering or false data injection on the device. 
\item \textit{Configurability:} 6thSense is configurable to provide different needs and flexible deployments. For example, 6thSense offers both online and offline training mode for different machine learning detection techniques to reduce power consumption. 
\end{itemize}

\vspace{-0.2cm}
\section{Detection Techniques: Theoretical Foundation}\label{sec:AnalyticalModel}
In this section, we describe the theoretical foundation of the detection techniques used in 6thSense. For the context-aware IDS in 6thSense, we utilize several different ML-based techniques including Markov Chain, Naive Bayes and, a set of other ML algorithms (e.g., PART, Logistic Function, J48, LMT, Hoeffding Tree, and Multilayer Perception) to differentiate between normal and malicious behavior on a smart device. \par
As explained in Section 4, we consider the context awareness of user activities in a smart device which shows state transition and sensor co-dependence feature in a smart device. The Markov Chain model can illustrate these properties of the smart device's sensors accurately based on different user activities in the transition matrix. Another advantage of using Markov Chain model is that it is easy to build the model from a large dataset and computational requirements are modest which can be met by resource-limited devices. On the other hand, the Naive Bayes model can build multiple activity contexts from sensor data and identifies whether a test dataset belongs to a user activity or a malicious activity. The Naive Bayes model uses the sensor co-dependence feature to build the activity context and classifies data accordingly. In addition to this, the Naive Bayes technique is chosen for its fast computation rate, small training dataset requirement, and ability to modify it with new training data without rebuilding the model from scratch. \par
Apart from the Markov Chain and the Naive Bayes model, other ML techniques are also common in malware detection because of their high accuracy rate~\cite{ye2017survey, sikder2019context3}. We also investigate how other ML algorithms perform in building a context-aware model from sensor data and detecting sensor-based threats on a smart device. Our main purpose is to check whether popular ML algorithms can understand and build an effective context-aware model for sensor-based threats. A discussion of these approaches in the context of 6thSense is given below. The efficacy of these different approaches utilized in 6thSense is analyzed in Section 7. 
\vspace{-10pt}
\subsection{Markov Chain-Based Detection}
\begin{figure}[tb]
  \centering
    \includegraphics[height=3.5cm, width=6cm]{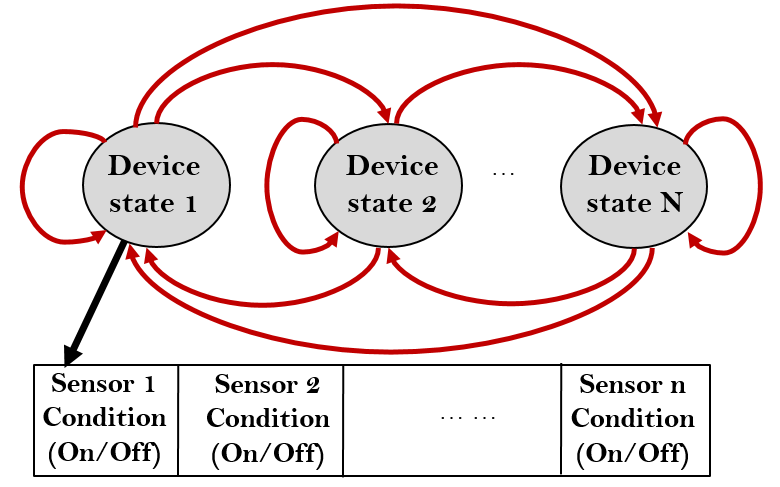}
    \vspace{-5pt}
      \caption{Markov Chain model for 6thSense.}
      \vspace{-10pt}
      \label{markov_fig}
  \vspace{-0.1cm}    
\end{figure}

Markov Chain model can be described as a discrete-time stochastic process which denotes a set of random variables and defines how these variables change over time. There are two main assumptions for Markov Chain model: (1) Probability distribution of the state at time \textit{t+1} depends on the state at time \textit{t} only. Here, the state refers to the overall condition of the stochastic process. (2) A state transition from previous timestamp (\textit{t}) to next timestamp (\textit{t+1}) is independent of time.
Markov Chain can be applied to illustrate a series of events where what state will occur next depends only on the previous state. In our study, a series of events represents user activity and state represents condition (i.e, values, on/off status) of the sensors in a smart device (Figure~\ref{markov_fig}). We can represent the probabilistic condition of Markov Chain as in Equation 1 where \textit{$X_t$} denotes the state at time \textit{t} \cite{keilson2012markov}.
\begin{equation}
\vspace{-2pt}
\footnotesize
\begin{split}
P(X_{t+1} = x| X_1 = x_1, X_2 = x_2 ..., X_t = x_t) = P(X_{t+1} = x| X_t = X_t) , \\ 
when,\ P(X_1 = x_1, X_2 = x_2 ..., X_t = x_t) > 0.
\end{split}
\vspace{-2pt}
\end{equation}

In our study, we observe the changes of condition of a set of sensors as a variable which changes over time. The condition of a sensor indicates whether the sensor value is changing or not from a previous value in time. Let us assume \textit{S} denotes a set which represents current conditions of \textit{n} number of sensors. So, \textit{S} can be represented with $S = \{S_1, S_2, S_3, ... , S_n\}$, where $S_1, S_2, S_3, ... , S_n = 0\ or\ 1$.
For a specific time, \textit{t}, we consider the combination of all the sensors' conditions in the smart device as the state of our model. As we consider change in a sensor's condition as binary output (1 or 0, where 1 denotes that sensor value is changing from previous instance and 0 denotes that sensor value is not changing), the number of total states of our model will be exponents of 2. For example, if we consider the total number of sensors in set \textit{S} is 10, the number of states in our Markov Chain will be $2^{10}$ and the states can be represented as a 10 bit binary number where each bit will represent the state of a corresponding sensor. For this, \textit{${p_{ij}}$} denotes the probability that the system in a state \textit{j} at time \textit{t+1} given that system is in state \textit{i} at time \textit{t}. If we have \textit{n} number of sensors  and \textit{${m=2^{n}}$} states in our model, Markov Chain can be constructed by the following transition probability matrix:
\begin{equation}
\footnotesize
P=\begin{bmatrix}
p_{11} & p_{12} & p_{13} & \ldots & \ldots & p_{1m}\\
p_{21} & p_{22} & p_{23} & \ldots & \ldots & p_{2m}\\
\ldots & \ldots & \ldots & \ldots & \ldots & \ldots \\
\ldots & \ldots & \ldots & \ldots & \ldots & \ldots \\
p_{m1} & p_{m2} & p_{m3} & \ldots & \ldots & p_{mm} \\
\end{bmatrix}
\end{equation}

The transition probability matrix of this Markov Chain can be constructed by observing the transitions from one state to another state for a certain time. Assume that, system's states are \textit{${X_0, X_1, \ldots, X_T}$} at a given time \textit{${t= 0, 1, \ldots, T}$}. We can represent the transition probability matrix as follows:
\begin{equation}
\footnotesize
{P_{ij}} = \frac{N_{ij}}{N_i},
\end{equation}
where,
\textit{${N_{ij}}$} = the number of transition from \textit{${X_t}$} to \textit{${X_{t+1}}$} where \textit{${X_t}$} in state \textit{i} and \textit{${X_{t+1}}$} in state \textit{j};  
\textit{${N_{i}}$} = the number of transition from \textit{${X_t}$} to \textit{${X_{t+1}}$}, where \textit{${X_t}$} in state \textit{i} and  \textit{${X_{t+1}}$} in any other state. The initial probability distribution of this Markov Chain can be represented as follows:
\begin{equation}
\footnotesize
Q = \begin{bmatrix}
q_1 & q_2 & q_3 & \ldots & \ldots & q_m
\end{bmatrix}
\end{equation}

where \textit{${q_m}$} is the probability that the model is in state \textit{m} at time 0. Probability of observing a sequence of states \textit{${X_1, X_2, \ldots, X_T}$} at a given time \textit{${1, \ldots, T}$} can be computed using the following equation:
\begin{equation}
\vspace{-0.2cm}
\footnotesize
P(X_1, X_2, \ldots, X_T) = {q_{x1}} \prod_{2}^{T} {P_{{X_{t-1}}X{t}}}
\end{equation}
For 6thSense, instead of predicting the next state, we determine the probability of occurring a transition between two states at a given time. We train our Markov Chain model with a training dataset collected from real users and build the transition matrix accordingly. Then, we determine sensor working condition for time \textit{t} and \textit{t+1}. Let us assume \textit{a} and \textit{b} are sensor's state in time \textit{t} and \textit{t+1}. We determine the probability of transition from state \textit{a} to \textit{b} which can be found by looking up in the transition matrix and calculating \textit{P(a,b)}. As the training dataset consisted of sensor data from benign activities, we can assume that if transition from state \textit{a} to \textit{b} is malicious, the calculated probability from transition matrix will be zero.
\vspace{-0.4cm}
\subsection{Naive Bayes Based Detection}
The Naive Bayes model is a simple probability estimation method which is based on Bayes' method. The main assumption of the Naive Bayes detection is that the presence of a particular sensor condition in a task/activity has no influence over the presence of any other feature on that particular event. The probability of each event can be calculated by observing the presence of a set of specific features.

Assume \textit{${p(x_1,x_2)}$} is the general probability distribution of two events \textit{${x_1, x_2}$}. Using the Bayes rule, we can have the following equation: 
\vspace{-0.1cm}
\begin{equation}
\footnotesize
\vspace{-0.1cm}
p(x_1,x_2) = p(x_1|x_2)p(x_2),
\end{equation}
where \textit{${p(x_1|x_2)}$} = Probability of the event \textit{${x_1}$} given that event \textit{${x_2}$} will happen. Now, if we have another variable, \textit{c}, we can rewrite Equation 7 as follows: 
\vspace{-0.1cm}
\begin{equation}
\vspace{-0.1cm}
p(x_1,x_2|c) = p(x_1|x_2,c)p(x_2|c).
\end{equation}
If knowledge of \textit{c} is sufficient enough to determine the probability of event \textit{${x_1}$}, we can state that there is conditional independence between \textit{${x_1}$} and \textit{${x_2}$}~\cite{mukherjee2012intrusion}. So, we can rewrite the first part of Equation 8 as \textit{${p(x_1|x_2,c) = p(x_1|c)}$}, which modifies Equation 8 as follows: 
\vspace{-0.1cm}
\begin{equation}
\vspace{-0.1cm}
\footnotesize
p(x_1,x_2|c) = p(x_1|c)p(x_2|c).
\end{equation}

\begin{figure}[t!]
  \centering

    \includegraphics[height=4cm, width=8cm]{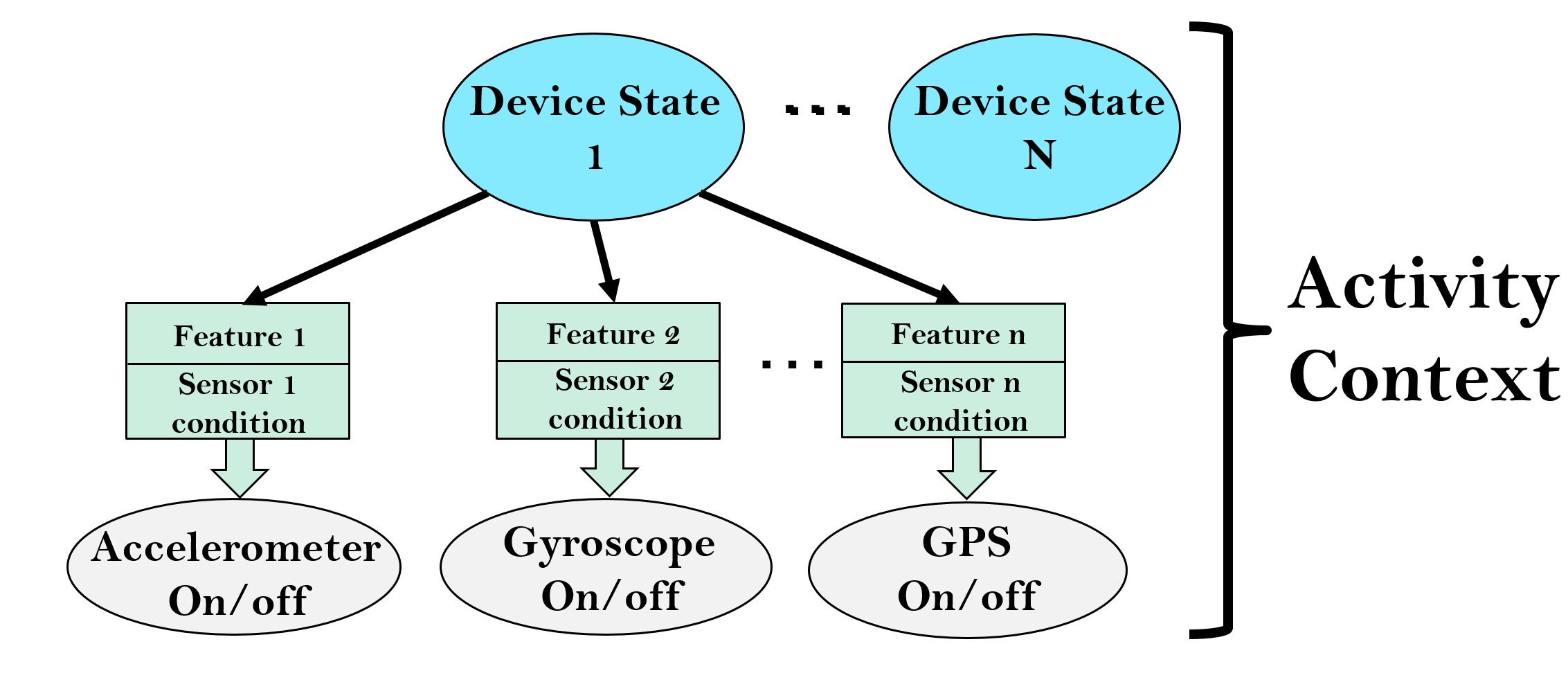}
    \vspace{-10pt}
      \caption{Naive Bayes model for 6thSense.}
      \vspace{-14pt}
      \label{naive_fig}
\end{figure}

In 6thSense, we consider users' activity as a combination of \textit{n} number of sensors (Figure~\ref{naive_fig}). Assume X is a set which represents current conditions of \textit{n} number of sensors. We consider that conditions of sensors are conditionally independent (See Section 4.2), which means a change in one sensor's working condition has no effect over a change in another sensor's working condition. As we explained earlier, the probability of executing a task depends on the conditions of a specific set of sensors. So, in summary, although one sensors' condition does not control another sensor's condition, overall probability depends on all the sensors' conditions. For example, if a person is walking with his smartphone in his hand, the motion sensors (accelerometer and gyroscope) will change. However, this change will not force the light sensor or the proximity sensor to change its condition. Thus, sensors in a smart device change their conditions independently, but execute a task together. From Equation 9, we can have a generalized formula for this context-aware model~\cite{mukherjee2012intrusion}: 
\begin{equation}
\footnotesize
p(X|c)=\prod_{i=1}^{n} p(X_i|c).
\end{equation}
In our contextual activity-oriented model, we have a set of training data for users' activities. Assume that \textit{B} represents a set which denotes \textit{m} numbers of user activities. We can determine the probability of a dataset X to be classified as a user activity using the following equation: 
\begin{equation}
\footnotesize
P(B_i|X) = \frac{P(X|B_i)P(B_i)}{P(X)},\
\end{equation}
where i = 1, 2, \ldots , m. As the sum of all the conditional probabilities for X will be 1, we can have the following equation which will lead to Equation 12---
\vspace{-0.1cm}
\begin{equation}
\footnotesize
\sum_{i=1}^m P(B_i|X) = 1.
\end{equation}
\vspace{-0.1cm}
\begin{equation}
\footnotesize
P(B_i|X) = \frac{P(X|B_i)P(B_i)}{\sum_{i=1}^m P(X|B_i) P(B_i)}.
\end{equation}

This calculated conditional probability then is used to determine the benign user activity or malicious attacks in 6thSense. In this way, we compute the probability of occurring an activity over a certain period of time.\par

We divide the sensor data into smaller time values (1 second) and calculate the probability of each instances to infer the user activity. The calculated probability per second data is then used in the expected value theorem to calculate total probability. If the probability of the first instance is \textit{${p_1}$} with a value of \textit{${a_1}$}, probability of the second instance is \textit{${p_2}$} with a value of \textit{${a_2}$} and so on, up to value \textit{${a_n}$}, the expected value can be calculated by the following formula:
\begin{equation}
\footnotesize
E[N]= \frac{a_1p_1+a_2p_2+a_3p_3+\ldots \\ \ldots+a_np_n}{a_1+a_2+\ldots \ \ldots+ a_n}.
\end{equation}

As all the values of \textit{${a_1}$}, \textit{${a_2}$}, ... ..., \textit{${a_n}$} are equally likely, this expected value becomes a simple average of cumulative probability of each instances. We infer the user activity by setting up a configurable threshold value in the 6thSense framework and checking whether calculated value is higher than the threshold or not. If it is lower than the threshold value, a malicious activity is occurring in the device. 
\vspace{-0.4cm}
\subsection{Other ML-based Detection Techniques}
In addition to the Markov Chain and the Naive Bayes models above, there are other machine learning algorithms (such as  PART, Logistic Function, J48, LMT, Hoeffding Tree, and Multilayer Perception) that are very popular for anomaly detection schemes because of their fast computation and easy implementation. 

In the alternative detection techniques, we used four types of ML-based classifier to build a context-aware analytical model for 6thSense. The following briefly discusses these classifiers and our rationale to include them: 
\par \textit{Rule-based Learning.} Rule-based ML works by identifying a set of relational rules between attributes of a given dataset and represents the model observed by the system~\cite{gu2007bothunter}. The main advantage of the rule-based learning is that it identifies a single model which can be applied commonly to any instances of the dataset to make a prediction of outcome. As we train 6thSense with different user activities, the rule-based learning provides one model to predict data for all the user activities which simplifies the framework. For 6thSense, we chose, \textit{PART} algorithm for the rule-based learning. 
\par\textit{Regression Model.} Regression model is widely used in data mining for its fast computation. This type of classifier observes the relations between dependent and independent variables to build a prediction model \cite{shabtai2012andromaly}. For 6thSense, we have a total 11 attributes where we have one dependent variable (device state: malicious/benign) and ten independent variables (sensor conditions). Regression model observes the change in the dependent variable by changing the values of the independent variables and build the prediction model. We use the logistic regression model in 6thSense,  which also yields with high accuracy against conventional Android malware~\cite{shabtai2012andromaly}. 
\par \textit{Neural Network.} Neural network is another common technique that is utilized by researchers for malware detection. In neural network techniques, the relation between attributes of dataset is compared with the biological neurons and a relation map is created to observe the changes for each attribute~\cite{linda2009neural}. We chose \textit{Multilayer Perceptron} algorithm for training the 6thSense framework as it can distinguish relationships among  non-linear dataset. 
\par \textit{Decision Tree.} Decision tree algorithms are predictive models where decision maps are created by observing the changes in one attribute in different instances \cite{ye2007imds}. These types of algorithms are mostly used in a prediction model where output can have a finite set of values. For 6thSense, we utilized and tested three different decision tree algorithms (\textit{J48}, \textit{LMT (Logistic Model Tree)}, and \textit{Hoeffding tree}) to compare the outcome of our framework.

\vspace{-10pt}
\section{6thSense Framework} \label{sec:Framework}
\begin{figure}[t!]
  \centering
  \framebox{
    \includegraphics[height=6cm, width=6.2cm]{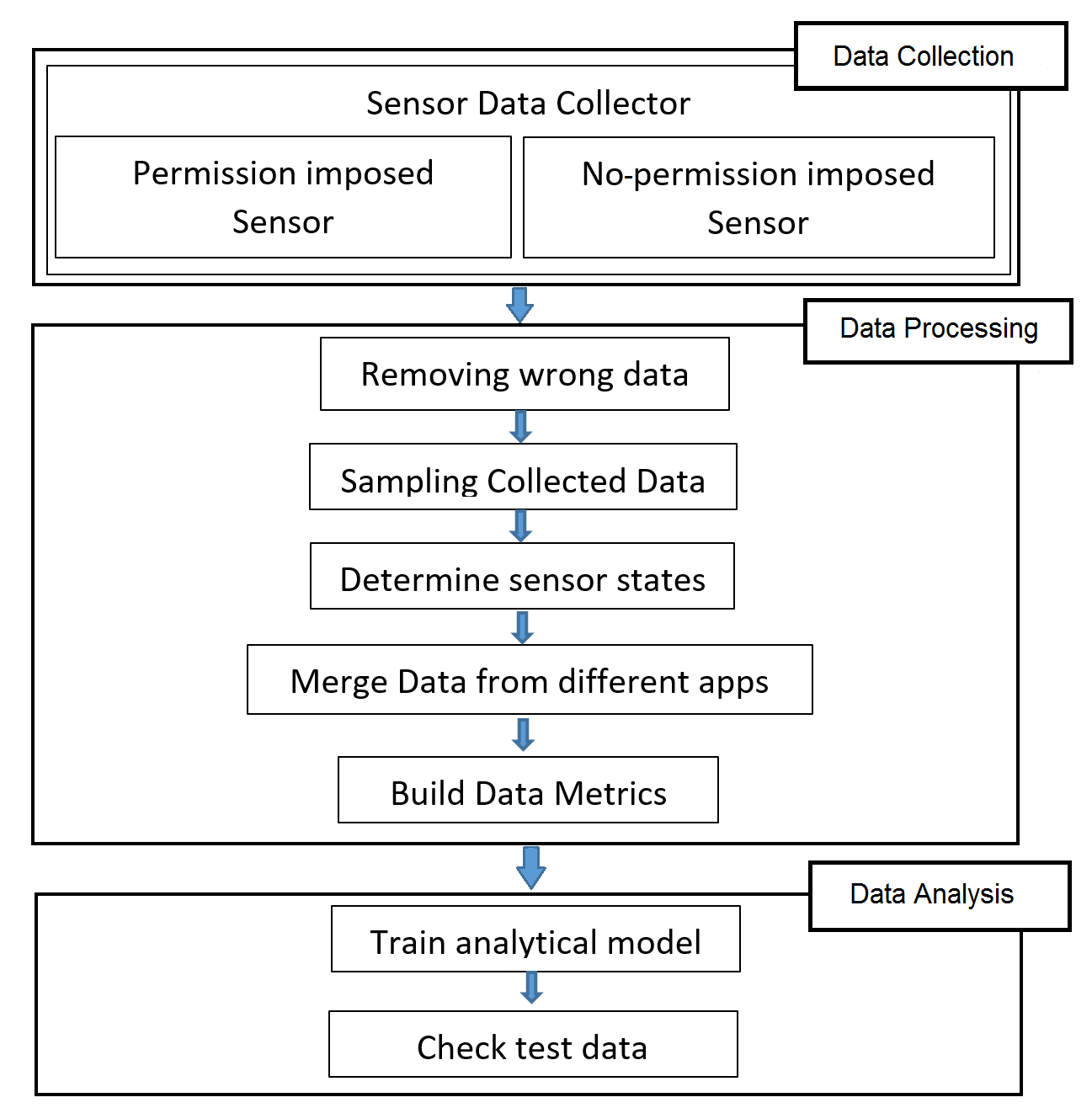}
    }
      \caption{Overview of 6thSense.}
      \vspace{-0.5cm}
      \label{overviewfig}
\end{figure}

\vspace{-2pt}
In this section, we provide a detailed overview of our proposed context-aware IDS framework, 6thSense, for detecting sensor-based threats on smart devices. As illustrated in Figure~\ref{overviewfig}, 6thSense has three main phases: (1) data collection, (2) data processing, and (3) data analysis. In the Data Collection phase, we use a custom Android App to collect the sensor data for different user activities and the collected sensor data are then processed in the Data Processing phase. In Phase 3, the collected data is fed into detection models and the end result indicates whether the current state of the device is malicious or not. The following sub-sections briefly describe these three phases.
\vspace{-10pt}
\subsection{Data Collection Phase}
In this phase, 6thSense collects data from different sensors of a smart device. There can be multiple sensors in a smart device. 6thSense considers nine sensors in total to identify different user activities using a sensor-rich Android device. The sensors selected are accelerometer, gyroscope, light sensor, proximity sensor, GPS, audio sensor (microphone and speaker), camers, and headphone. The chosen sensors are then categorized into two following categories.

\textit{No-permission-imposed sensors in 6thSense:}  For 6thSense, we chose four no-permission imposed sensors (i.e., accelerometer, gyroscope, light, proximity sensors). We can also refer these sensors as data-oriented sensors in the context of 6thSense because values provided by these sensors need to be observed to infer user activities. For example, accelerometer's and gyroscope's values change with motion and they give values on \textit{X, Y,} and \textit{Z} axes. To detect whether a sensor is activated or not for a specific activity, one needs to observe values of these sensors. 

\textit{Permission-imposed sensors in 6thSense:} We chose five permission-imposed sensors to build the context-aware model (microphone, GPS, speaker, camera, and headset) of 6thSense. The conditions of these sensors can be represented by their logical states (on/off status) for different user activities. Hence, we also referred to these sensors as logic-oriented sensors in the context of 6thSense. For example, microphone has only two values to identify users' activity: on and off. So, it can be represented with 0 or 1 to detect if the camera is on or off correspondingly.  

\par To collect the data and logical values from sensors, we built a custom Android App and 6thSense used this in the data collection phase. In Android, this App uses \textit{sensoreventlistener} API to log numerical values of the data-oriented sensors. On the other hand, the App determines the state of the sensor and logs 0 or 1 if the sensor is on or off, respectively. This App uses the user permission to use the microphone, GPS, and camera to record the working conditions of these sensors. For GPS, we consider two datasets - either GPS is turned on or not and either location is changing or not. In total, six different logic state information for five aforementioned permission-imposed sensors are collected by this App.

Note that 6thSense considers different typical daily human activities~\cite{activity4} that involve the smart devices (e.g., smart watch, smart phone, etc.) to build the contextual model. These activities include walking, talking, interacting (playing games, browsing), driving (as driver and passenger). Furthermore, the number of activities is configurable in 6thSense and is not limited to aforementioned examples. As also explained in the evaluation of 6thSense, a total of seven and nine typical daily activities are selected for smart watch and smart phone respectively as they are considered as common user activities~\cite{activity}. 6thSense collects these data using the App for different users to train the framework which is then used to distinguish the normal sensor behavior from the malicious behavior. 
In summary, the aforementioned App collects data from eight different sensors for different typical user activities. 6thSense observes sensor state (combination of working conditions (i.e., values, on/off status) of different sensors) in a per second manner for each user activity. Each second of data for user activity corresponds to 512 state information from eight different sensors. 
\vspace{-0.2cm}
\subsection{Data Processing Phase}
In the second phase of the framework, 6thSense organizes the data to use. As different sensors have different frequencies on the smart device, the total number of readings of sensors for a specific time period is different. For example, the accelerometer and gyroscope of \textit{LG Watch Sport} have a sampling frequency of approximately 418 Hz and 32 Hz, respectively while the light sensor has a sampling frequency of 5 Hz. Thus, the data collected in Phase 1 needs to be sampled and reorganized. 6thSense observes the change in the sensor condition in each second to determine the overall state of our device and from this per second change, 6thSense determines the activity of users. For this reason, 6thSense takes all the data given by a single sensor in a second and calculates the average value of the sensor reading. This process is only applicable for the data oriented sensors as mentioned earlier. Again, the data collected from the App is numerical value as given by the sensor. However, for the detection model, 6thSense only considers the condition of the sensors. 6thSense observes the data collected by the aforementioned App and determines whether the condition of sensors is changing or not. If the sensor value is changing from the previous value in time, 6thSense represents the sensor condition as 1 and 0 otherwise. The logic state information collected from the sensors need to be reorganized, too as these data are merged with the data collected from the collected values from the other sensors to create an input matrix. We consider the condition of the sensors to be the same over time and organize the data accordingly. The reorganized data generated from the aforementioned App are then merged to create the training matrices.
\vspace{-0.4cm}
\subsection{Data Analysis Phase}
\vspace{-0.1cm}
In the third, 6thSense uses different ML-based detection techniques introduced in the  previous section to analyze the data matrices generated in the previous phase. 

\par For the Markov Chain-based detection, 6thSense uses 75\% of the collected data to train 6thSense and generate the transition matrix. This transition matrix is used to determine whether the transition from one state to another is appropriate or not. Here, state refers to generic representation of all the sensors' conditions on a device. For testing purposes, we have two different data set --- benign activities or trusted model and malicious activities or threat model. The trusted model consists of 25\% of the collected data for different user activities. We tested the trusted model to ensure the accuracy of the 6thSense framework in detecting benign activities. The malicious activities are built from performing the attack scenarios mentioned in Section~\ref{sec:Adversarymodel}. 6thSense calculates the probability of a transition occurring between two states at a given time and accumulates the total probability to distinguish between normal and malicious activities. 

\par To implement the Naive Bayes-based detection technique, 6thSense uses the training sessions to define different user activities. In 6thSense, seven typical user activities are selected in total for smart watch as listed in Table 3. In addition to these user activities, we consider walking with smart device in pocket and making a video call as typical user activities to test 6thSense in smart phone. 6thSense uses ground truth user data to define these activities. Using the theoretical foundation explained in Section~\ref{sec:AnalyticalModel}, 6thSense calculates the probability of a test session to belong to any of these defined activities. As 6thSense considers one second of data in each computational cycle, the total probability up to a predefined configurable time interval (in this case five minutes) is calculated. This calculated probability is used to detect malicious activities from normal activities. If the computed probability for all the known benign activities is not over a predefined threshold, then it is detected as a malicious activity. 

\par For the other alternative machine-learning-based detection techniques, 6thSense uses WEKA, a data mining tool which offers data analysis using different machine learning approaches~\cite{hall2009weka}.  

\vspace{-10pt}
\section{Performance Evaluation of 6thSense} \label{sec:Evaluation}
\begin{table*}[htb!]
\centering
\fontsize{8}{10}\selectfont
\resizebox{0.8\textwidth}{!}{
\begin{tabular}{cccc}
\toprule
\textbf{\begin{tabular}[c]{@{}c@{}}Sensor \\ type\end{tabular}} & \textbf{Name} & \textbf{\begin{tabular}[c]{@{}c@{}}Model \\(Smart Watch $\mid$ Smart Phone)\end{tabular}} & \textbf{\begin{tabular}[c]{@{}c@{}}Specification \\ (Smart Watch $\mid$ Smart Phone)\end{tabular}} \\ \hline
\midrule
\multirow{4}{*}{\begin{tabular}[c]{@{}c@{}}No-permission imposed \\sensors\end{tabular}}  & Accelerometer  & \begin{tabular}[c]{@{}c@{}}Bosch BMI160 Acceleration Sensor $\mid$\\MPU6500 Acceleration Sensor\end{tabular}   & \begin{tabular}[c]{@{}c@{}}78.4532 $m/s^2$, 417.67 Hz, 0.01 mA $\mid$\\ 19.6133 $m/s^2$, 203.60 Hz, 0.25 mA\end{tabular} \\ \cline{2-4} 
                         & Gyroscope      & \begin{tabular}[c]{@{}c@{}}Bosch BMI160 Gyroscope Sensor $\mid$ \\ MPU6500 Gyroscope Sensor\end{tabular}     & \begin{tabular}[c]{@{}c@{}}17.453293 rad/s, 31.95 Hz, 0.01 mA $\mid$ \\ 8.726646 rad/s, 203.60 Hz, 6.1 mA \end{tabular}\\\cline{2-4} 
                         & Light Sensor   & \begin{tabular}[c]{@{}c@{}}APDS-9306 Light Sensor $\mid$ \\ TMG399X RGB Sensor  \end{tabular}   	& \begin{tabular}[c]{@{}c@{}}30000 lux, 5 Hz, 0.11 mA $\mid$ \\600000 lux, 5.62 Hz, 0.75 mA  \end{tabular} \\\cline{2-4}
                         & Proximity Sensor  & \begin{tabular}[c]{@{}c@{}}LG Wear Detection Sensor $\mid$ \\ TMG399X proximity sensor \end{tabular}&  \begin{tabular}[c]{@{}c@{}}1V, 0.15 mA $\mid$ 8V, 0.75 mA\end{tabular} \\\cline{2-4}
                         \hline
\multirow{3}{*}{\begin{tabular}[c]{@{}c@{}}Permission-imposed sensors\end{tabular}}  
                          & Microphone  &  \begin{tabular}[c]{@{}c@{}}Qualcomm Snapdragon Wear 2100\\ built in microphone $\mid$ Qualcomm \\Snapdragon 801 Processor built \\in microphone\end{tabular}   & \begin{tabular}[c]{@{}c@{}}120 dB, .12 mA $\mid$ 86 dB, .75 mA \end{tabular}        \\ \cline{2-4}
                          & Speaker     &  \begin{tabular}[c]{@{}c@{}}Qualcomm Snapdragon Wear 2100\\ built in speaker $\mid$ Qualcomm \\Snapdragon 801 Processor built\\ in speaker\end{tabular}   & \begin{tabular}[c]{@{}c@{}}90 dB, .18 mA  $\mid$ 110 dB, 1 mA  \end{tabular}  \\ \cline{2-4}
                          & Camera & N/A $\mid$ Samsung S5K2P2XX & N/A $\mid$ 12 megapixels, 30 fps, 4.7 mA\\ \hline
\bottomrule                                                                          
\end{tabular}}
\caption{Sensor list of LG Watch Sport and Samsung Galaxy S5 Duos smartphone used in experiment.}
\vspace{-0.6cm}
\label{tab:sensors}
\end{table*}
In this section, we evaluate the efficiency of the proposed context-aware IDS framework, 6thSense, in detecting the sensor-based threats on smart devices (smartphone and smart watch). We tested 6thSense with the data collected from different users for benign activities and adversary models described in Section~\ref{sec:Adversarymodel}. As discussed earlier, 6thSense considers three sensor-based threats: (1) a malicious App that can be triggered via a light or motion sensors, (2) a malicious App that can leak information via audio sensors, and (3) a malicious App that steals data via audio sensors. 
Furthermore, we measured the performance impact of 6thSense on the devices 
and present a detailed results for the efficiency of the 6thSense framework on both a smart watch and smart phone.
\vspace{-0.4cm}
\subsection{Training Environment and Dataset} \label{sec:trainingenv}
In order to test the effectiveness of 6thSense, we implemented it on both a sensor-rich Android-based smart watch and smartphone. We used the \textit{LG Watch Sport} as a reference Android smart watch with \textit{Android Wear version 2.0} to collect sensor data for different typical user activities. We chose this Android device as  the LG watch sport is a second generation, stand-alone Android wearable that provides a rich set of sensors\modificationend. A list of sensors of \textit{LG Watch Sport} is given in Table~\ref{tab:sensors}. As discussed earlier, we selected 7 different typical user activities or tasks to collect user data (Table 2). These are typical basic activities with the smart watches that people usually do in their daily lives~\cite{activity}. The user activities/tasks are categorized in two categories as generic activities and user related activities.

\par \textit{Generic activities} are the activities in which the sensor readings are not affected by the smart device users. Sleeping wearing smart watch, driving with the smart watch using GPS as a navigator, and driving with smart watch in hand are three generic activities that were considered in this work. Basically, in the generic activities, sensors' data are not affected by different users since users do not interact with the smart watch directly. For example, if a user is sleeping, sensors activity will be irregular depending on sleeping pattern. There will be less movement detected in the device and sensor data will be changed accordingly. For \textit{user-related activities}, in which the sensor readings may be affected by the user, we identified four different activities including walking, playing games, browsing, and making voice calls via smart watch. 

For implementing and evaluating the performance of 6thSense on smartphone, we chose \textit{Samsung Galaxy S5 Duos} with Android OS version 7.1.2 (Android N) which provides a broad range of sensors. Samsung currently holds approximately 23.3\% of total market share of smartphones~\cite{Samsung} and is the largest Android operated smartphone manufacturer which motivates to implement 6thSense on Samsung smartphone. In addition to user activities used in the smart watch data collection, we considered two more user-related activities (walking with the device in pocket/bag and making video calls) for testing 6thSense on the smartphone. 

6thSense was tested by 100 different individuals (50 smart watch users and 50 smartphone users) while the sensor data was collected from the smart watch and the smartphone. We note that our study with human subjects was approved by the appropriate Institutional Review Board (IRB) and we followed all the procedures strictly in our study. To train and test 6thSense on the smart watch, we collected 200 sets of data for four user-related activities for the smart watch where each dataset comprised of 300 seconds of data from the selected sensors mentioned in Section~\ref{sec:Framework}. We also collected three sets of data for each general activity. We asked the different users to perform the same activity to ensure the integrity for different tasks. We also asked the users to perform the tasks naturally without any influence of the lab environment. Users performed these tasks in a real-life workplace and outdoor in a natural environment. Additionally, users chose their preferred place, walking routes, and apps in the entire data collection process. For example, to collect data in walking scenario, users chose their preferred walking routes both inside their workplace and outside environment. Note that each five minutes of the data collected for user-related and generic activities corresponds to 300 events with 512 different states. So, a total of 153,600 different event-state information were analyzed by 6thSense for a user activity. For testing 6thSense on a smartphone, we collected data from 50 different individuals for nine different activities. We considered nine different sensors to build the context-aware model and each dataset depicted 300 events with 1024 different states and a total of 307,200 event-state information~\cite{203854}. \par
\vspace{-2pt}
For the malicious dataset, we created three different attack scenarios considering the adversary model mentioned in Section 4. For Threat 1, we developed two different Android Apps which could be triggered using the light sensor and motion sensors on the smart watch. We also created the same malicious Android app for the smart phone. To perform the attack described in Threat 2, we developed a malware that could record conversations as audio clips and playback after a specific time to leak the information. This attack scenario included both the microphone and speaker on the smart watch and smart phone. We developed another version of this app which could record conversations as audio clips in smartphone using a connected smart watch. Also, for Threat 3, we developed a malicious App that could scan all the sensors and if none of the sensors were changing their working conditions, the malicious App could open up the microphone and record audio clips surreptitiously. For Threat 3, we developed another version for smart devices with camera (e.g., smartphone) where a malicious App can scan all the sensors of a device and if device was inactive, the malicious App could activate camera and record videos covertly. We developed an updated version of this attack which could start recording via microphone in a smart watch if the connected smartphone was inactive. This version of the app could bypass the security feature introduced on Android P~\cite{andrp}. In summary, we created 10 different malware that could perform malicious activities in Android-powered smart phone and smart watch. We collected 18 different datasets (a total of 62,850 event-state information) from these three attack scenarios to test the efficacy of 6thSense against these adversaries in a smart watch. \par

In order to test 6thSense, we divided the collected real user data into two sections as it is a common practice~\cite{Radh1310:Passive}. 75\% of the  collected benign dataset was used to train the 6thSense framework and 25\% of the collected data along with malicious dataset were used for testing purposes. For the Markov Chain-based detection technique, the training dataset was used to compute the state transitions and to build the transition matrix. On the other hand, in the Naive Bayes-based detection technique, the training dataset was used to determine the frequency of sensor condition changes for a particular activity or task. As noted earlier, for the smart watch, there were seven activities for the Naive Bayes technique. We split the data according to their activities for this approach. 
\begin{table}[h!]
\vspace{-4pt}
\centering
\resizebox{0.35\textwidth}{!}{
\begin{threeparttable}
\centering
\begin{tabular}{p{3cm}p{3.7cm}}
\toprule
\textbf{Task Category}                            & \textbf{Task Name}   \\ \hline
\midrule
\multirow{3}{*}{Generic Activities}      & 1. Sleeping                         \\ \cline{2-2} 
                                         & 2. Driving as driver                \\ \cline{2-2} 
                                         & 3. Driving as passenger             \\ \hline
\multirow{6}{*}{User-related Activities} & 1. Walking with smart watch in hand       \\ \cline{2-2} 
                                         & 2. Playing games                    \\ \cline{2-2} 
                                         & 3. Browsing                         \\ \cline{2-2}
                                         & 4. Making phone calls                  \\\cline{2-2} 
                                         & 5. Walking with device in pocket/bag\tnote{\textdagger}                  \\\cline{2-2}
                                         & 6. Making video calls\tnote{\textdagger}                  \\\bottomrule
\end{tabular}
\begin{tablenotes}
    \item[\textdagger] Only considered for smart phone.
\end{tablenotes}
\label{tasks}
\vspace{-0.2cm}
\end{threeparttable}}

\caption{Typical activities of users on a smart device~\cite{activity}.}
\vspace{-0.25cm}
\end{table}
For the analysis of the other ML-based approaches, the data in benign and malicious classes were used to train and test 6thSense using 10-fold cross validation for different ML algorithms.
\begin{table*}[htb!]
\centering
\resizebox{0.8\textwidth}{!}{
\begin{threeparttable}
\begin{tabular}{ccccccccccccc}
& \multicolumn{6}{c}{\textbf{Smart Watch}} & \multicolumn{6}{c}{\textbf{Smart Phone}}\\ \hline
\multicolumn{1}{|c|}{\textbf{Threshold}\tnote{\textdagger}} & \textbf{\begin{tabular}[c]{@{}c@{}}Recall\end{tabular}} & \textbf{\begin{tabular}[c]{@{}c@{}}FN\end{tabular}} & \textbf{\begin{tabular}[c]{@{}c@{}}Precision\end{tabular}}   & \textbf{\begin{tabular}[c]{@{}c@{}}FP\end{tabular}} & \textbf{Accuracy} & \textbf{F-score} & \multicolumn{1}{|c}{\textbf{Recall}} & \textbf{\begin{tabular}[c]{@{}c@{}}FN\end{tabular}} & \textbf{\begin{tabular}[c]{@{}c@{}}Precision\end{tabular}}   & \textbf{\begin{tabular}[c]{@{}c@{}}FP\end{tabular}} & \textbf{Accuracy} & \multicolumn{1}{c|}{\textbf{F-score}} \\ \hline

0   &  0.66  &  0.33  &  1  &  0      &  0.77 &  0.79 & 0.62 & 0.38 & 1 & 0 & 0.68  & 0.76\\ 
1   &  0.77  &  0.22  &  1  &  0      &  0.85 &  0.87 & 0.86 & 0.14 & 1 & 0 & 0.88  & 0.92 \\ 
\rowcolor{Gray}
2   &  0.88  & 0.11   & 1   &  0      & 0.92  & 0.94  & 0.96 & 0.04 & 1 & 0 & 0.96  & 0.97\\ 
\rowcolor{Gray}
3   &  0.97  & 0.02   & 0.98 & 0.01   & 0.97  & 0.98  & 0.98 & 0.02 & 1 & 0 & 0.98  & 0.98\\ 
\rowcolor{Gray}
5   &  0.99  & 0.001  & 0.89  & 0.11  & 0.96  & 0.94  & 1    & 0    & 0.9   & 0.1   & 0.98   & 0.94\\ 
6   &  1     & 0      & 0.84  & 0.16  & 0.95  & 0.92  & 1    & 0    & 0.8   & 0.2   & 0.96   & 0.89 \\ 
\hline
\end{tabular}
\begin{tablenotes}
    \item[\textdagger] Number of consecutive malicious state is considered as threshold
\end{tablenotes}
\end{threeparttable}}
\caption{Performance evaluation of Markov Chain based model.}
\label{markovchain}
\vspace{-20pt}
\end{table*}
\vspace{-0.4cm}
\subsection{Performance Metrics}
In the evaluation of 6thSense, we utilized the following six different performance metrics: Recall rate (sensitivity or True Positive rate), False Negative rate, Specificity (True Negative rate), False Positive rate, Accuracy, and F-score. True Positive (TP) indicates number of benign activities that are detected correctly while true negative (TN) refers to the number of correctly detected malicious activities. On the other hand, False Positive (FP) states malicious activities that are detected as benign activities and False Negative (FN) defines number of benign activities that are categorized as malicious activity. F-score is the performance metric of a framework that reflects the accuracy of the framework by considering the recall rate and specificity. These performance metrics are defined as follows:

\begin{equation}
\footnotesize
{Recall\ rate\ (TP\ Rate)} = \frac{TP}{TP+FN} ,
\end{equation}
\begin{equation}
\footnotesize
{False\ negative\ rate} = \frac{FN}{TP+FN} ,
\end{equation}
\begin{equation}
\footnotesize
{Precision\ rate\ (TN\ rate)} = \frac{TN}{TN+FP} ,
\end{equation}
\begin{equation}
\footnotesize
{False\ positive\ rate} = \frac{FP}{TN+FP} ,
\end{equation}
\begin{equation}
\footnotesize
{Accuracy} = \frac{TP+TN}{TP+TN+FP+FN} ,
\end{equation}
\begin{equation}
\footnotesize
{F-score} = \frac{2*Recall\ rate*Precision\ rate}{Recall\ rate+Precision\ rate}. 
\end{equation}
In addition to the aforementioned performance metrics, we considered Receiver Operating Characterstic (ROC) curve and Precision Recall Curve (PRC) as other performance metrics for 6thSense. As our collected dataset is imbalanced (number of benign events is higher than the malicious events), the accuracy of the framework can be influenced by the dataset. To address data imbalance problem in 6thSense, we used PRC as a performance metric which considers data imbalance and reflects the base-rate fallacy correctly~\cite{Roy:2015:ESR:2818000.2818038}. 
\begin{figure*}[b!]
\vspace{-0.6cm}
\centering
\subfloat[ROC curve for Markov Chain]{\includegraphics[width=0.25\textwidth]{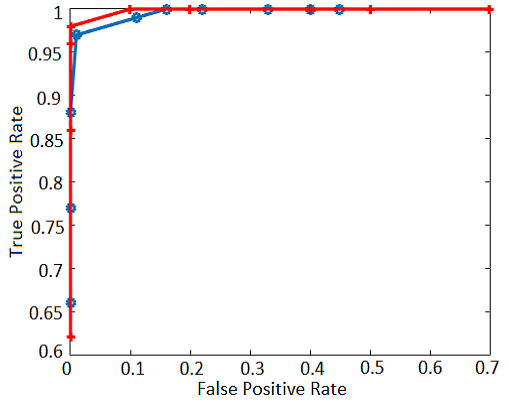}\label{fig:f1}}
\subfloat[PRC curve for Markov Chain]{\includegraphics[width=0.25\textwidth]{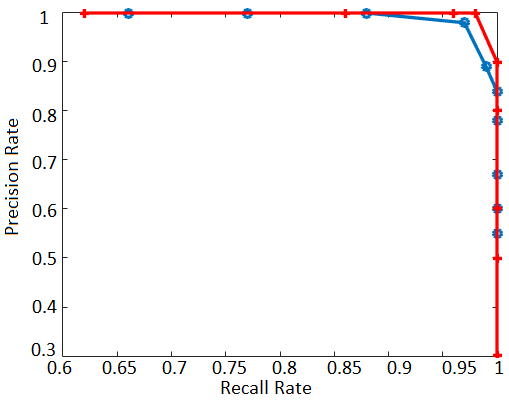}\label{fig:f2}}
\subfloat[ROC curve for Naive Bayes]{\includegraphics[width=0.25\textwidth]{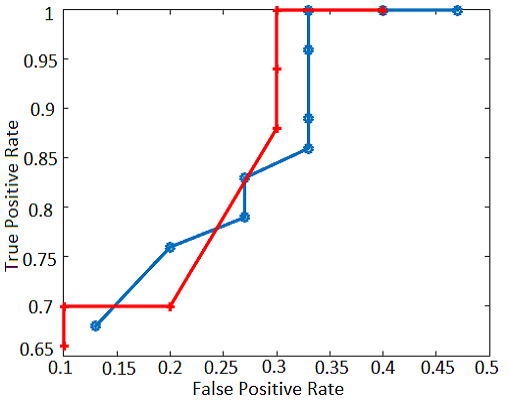}\label{fig:f3}}
\subfloat[PRC curve for Naive Bayes]{\includegraphics[width=0.25\textwidth]{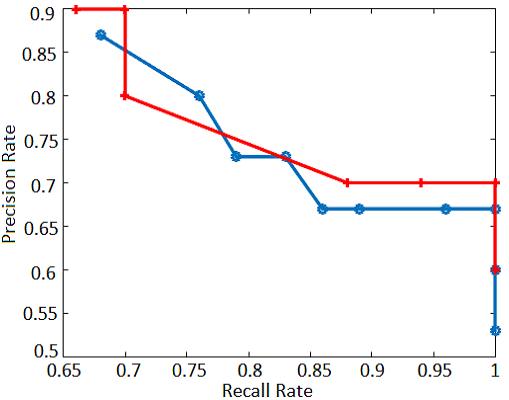}\label{fig:f4}}
\vspace{-2pt}
\caption{ROC curve and PRC curve of different detection techniques on smart watch (\modification \large-----\small\modificationend) and smart phone (\modif \large-----\small\modifend).}
      \label{fig}
\end{figure*}
\vspace{-6pt}
\subsection{Evaluation of Markov Chain-Based Detection}
In the Markov Chain-based detection technique, we question whether the transition between two states (sensors' on/off condition in each second) is expected or not. In the evaluation, we used 66 testing sessions in total for the smart watch, among which 51 sessions were for the benign activities (both generic and user-related activities) and the rest of the sessions were for the malicious activities. For evaluation in the smartphone, we have 80 testing sessions in total (65 benign sessions). A session is composed of a series of sensory context conditions where a sensory context condition is the set of all available sensor conditions (on/off state) for different sensors. As discussed earlier in Section 6, a sensor condition is a value indicating whether the sensor data is changing or not. In this evaluation, the sensory context conditions were computed every one second. For Markov Chain-based detection, we referred each sensory context condition as state of the device of that particular moment. 6thSense provides both online and offline training method to reduce performance overhead of the resource-constrained devices. As the highest battery life is 430 mAh for LG watch sport, training with different user data will consume more power which will increase power-accuracy trade-off of our framework; hence, we chose offline training method~\cite{LG}. For the test dataset, we used the transition matrix generated from the training period to determine whether transition from one state to another is malicious or not. We observed that in real devices, sometimes some sensor readings would be missed or real data would not be reflected due to hardware or software imperfections. And, real malicious Apps would cause consecutive malicious states on the device. Therefore, to overcome this, we also kept track of number of consecutive malicious states and used it as a threshold after which the session was considered as malicious. Table~\ref{markovchain} displays the evaluation results associated with the Markov Chain-based detection technique. When the threshold for consecutive malicious states is 0, i.e., when no threshold is applied, the accuracy is just 77\% and FNR is as high as 33\%. With increasing the threshold value, the accuracy first increases up to 97\% then starts decreasing. \par

The logical cut-off threshold should be three consecutive malicious occurrences which has both accuracy and F-score over 97\%. In Table~\ref{markovchain}, different performance indicators for Markov Chain based detection are also presented. We can observe that FN and TN rates of Markov Chain-based detection decrease as the threshold of consecutive malicious states increases. Again, both accuracy and F-score reach to a peak value with the threshold of three consecutive malicious states on the device. From Figure~\ref{fig}, we can see that FP rate remains zero while TP rate increases at the beginning. The highest TP rate without introducing any FP case is over 88\%. After 88\%, it introduces some FP cases in the system. For the cut-off threshold of three consecutive malicious occurrences, TP rate of 6thSense increases over 97\% with FP rates as low as 0.01\%. \par

Table~\ref{markovchain} also depicts evaluation of Markov chain model on the smartphone. Similar to the smart watch, TP rate and FP rate increase with consecutive malicious occurrences and FN and TN decrease with the threshold on a smartphone. The plausible cut-off threshold should be three consecutive malicious occurrences which is the same for the smart watch. The peak accuracy and F-score can be achieved for this threshold value which is over 98\%. From Figure~\ref{fig}, we can also notice that the highest possible TP rate without introducing any FP cases is 98\%. Figure~\ref{fig:f2} shows PRC curve for Markov Chain-based detection on both the smartwatch and the smart phone. We can see that for both the smart watch and the smartphone, area under PRC are approximately 1 which is ideal result for our imbalanced dataset.  In summary, Markov Chain-based detection in 6thSense can acquire accuracy over 97\% and auPRC approximately 1 with low FP rates (1.43\%) for both the smart watch and the smartphone.
\vspace{-0.5cm}
\begin{table*}[t!]
\centering
\resizebox{0.8\textwidth}{!}{
\begin{threeparttable}
\begin{tabular}{ccccccccccccc}
& \multicolumn{6}{c}{\textbf{Smart Watch}} & \multicolumn{6}{c}{\textbf{Smart Phone}}\\ \hline
\multicolumn{1}{|c|}{\textbf{Threshold}\tnote{\textdagger}} & \textbf{\begin{tabular}[c]{@{}c@{}}Recall\end{tabular}} & \textbf{\begin{tabular}[c]{@{}c@{}}FN\end{tabular}} & \textbf{\begin{tabular}[c]{@{}c@{}}Precision\end{tabular}}   & \textbf{\begin{tabular}[c]{@{}c@{}}FP\end{tabular}} & \textbf{Accuracy} & \textbf{F-score} & \multicolumn{1}{|c}{\textbf{Recall}} & \textbf{\begin{tabular}[c]{@{}c@{}}FN\end{tabular}} & \textbf{\begin{tabular}[c]{@{}c@{}}Precision\end{tabular}}   & \textbf{\begin{tabular}[c]{@{}c@{}}FP\end{tabular}} & \textbf{Accuracy} & \multicolumn{1}{c|}{\textbf{F-score}} \\ \hline

55\%    & 1 & 0 & 0.53  & 0.47 & 0.96  & 0.69 & 1  & 0  & 0.6  & 0.4 & 0.93  & 0.75    \\ 
57\%    & 1 & 0 & 0.6   & 0.4  & 0.96  & 0.75 & 1  & 0  & 0.7  & 0.3 & 0.95  & 0.82  \\ 
\rowcolor{Gray}
60\%    & 1 & 0 & 0.67  & 0.33 & 0.97  & 0.80 & 1  & 0  & 0.7  & 0.3 & 0.95  & 0.82 \\ 
\rowcolor{Gray}
62\%    & 0.96  & 0.04  & 0.67 & 0.33  & 0.94 & 0.79 & 1 & 0 & 0.7 & 0.3 & 0.95 & 0.82   \\ 
\rowcolor{Gray}
65\%    & 0.89  & 0.11  & 0.67 & 0.33  & 0.87 & 0.76 & 0.94  & 0.06 & 0.7 & 0.3 & 0.9 & 0.80   \\ 
67\%   & 0.86   & 0.14  & 0.67 & 0.33  & 0.85 & 0.75 & 0.88  & 0.12 & 0.7 & 0.3 & 0.85 & 0.78  \\ \hline
\end{tabular}
\begin{tablenotes}
    \item[\textdagger] Calculated expected probability is considered as threshold.
\end{tablenotes}
\end{threeparttable}}
\caption{Performance evaluation of Naive Bayes model.}
\label{naivetable}
\vspace{-5pt}
\end{table*}
\begin{table*}[t!]
\centering
\resizebox{0.8\textwidth}{!}{
\begin{tabular}{ccccccccccccc}
& \multicolumn{6}{c}{\textbf{Smart Watch}} & \multicolumn{6}{c}{\textbf{Smart Phone}}\\ \hline
\multicolumn{1}{|c|}{\textbf{Algorithms}} & \textbf{\begin{tabular}[c]{@{}c@{}}Recall\end{tabular}} & \textbf{\begin{tabular}[c]{@{}c@{}}FN\end{tabular}} & \textbf{\begin{tabular}[c]{@{}c@{}}Precision\end{tabular}}   & \textbf{\begin{tabular}[c]{@{}c@{}}FP\end{tabular}} & \textbf{Accuracy} & \textbf{F-score} & \multicolumn{1}{|c}{\textbf{Recall}} & \textbf{\begin{tabular}[c]{@{}c@{}}FN\end{tabular}} & \textbf{\begin{tabular}[c]{@{}c@{}}Precision\end{tabular}}   & \textbf{\begin{tabular}[c]{@{}c@{}}FP\end{tabular}} & \textbf{Accuracy} & \multicolumn{1}{c|}{\textbf{F-score}} \\ \hline

PART    & 0.98 & 0.012  & 0.69  & 0.30  & 0.98  & 0.80  & 0.99 & 0.01 & 0.65 & 0.35 & 0.99 & 0.79   \\ Logistic Function    & 0.99  & 0.01 & 0.35 & 0.65  & 0.97  & 0.49 & 0.99  & 0.01  & 0.28  & 0.72   & 0.99   & 0.43  \\ 
J48  & 0.99 & 0.01  & 0.71  & 0.29  & 0.99  & 0.81 & 0.99 & 0.01 & 0.65 & 0.35  & 0.99  & 0.79  \\ 
\rowcolor{Gray}
LMT     & 0.99 & 0.01  & 0.95   & 0.05 & 0.99  & 0.97 & 0.99 & 0.01 & 0.93  & 0.07 & 0.99  & 0.96   \\ 
Hoeffding Tree   & 1     & 0    & 0.07   & 0.93 & 0.99   & 0.12  & 1 & 0 & 0.06    & 0.94  & 0.99     & 0.11 \\ 
Multi-layer Perceptron    & 0.99 & 0.01   & 0.65 & 0.35   & 0.98  & 0.81 & 0.99 & 0.01  & 0.69  & 0.31   & 0.99  & 0.82   \\ \hline
\end{tabular}}
\vspace{-5pt}
\caption{Performance of other different machine learning based-detection techniques tested in 6thSense.}
\vspace{-15pt}
\label{wekatable}
\end{table*}
\subsection{Evaluation of Naive Bayes-based Detection}
In the Naive Bayes-based detection technique, 6thSense calculates the probability of a session to match it with each activity defined in Section~\ref{sec:trainingenv}. Here, 6thSense checks the calculated probability of an activity from dataset against a threshold to determine the correct activity. If there is no match for a certain sensor condition with any of the activities, 6thSense detects the session as malicious. Table~\ref{naivetable} shows the evaluation results. 

For the smart watch, for a threshold value of 55\%, FN rate is zero. However, FPR is too high (47\%), which lowers F-score of the framework. For a threshold of 60\%, FPR decreases while FNR is still zero. In this case, accuracy is 97\% and F-score is 80\%. If the threshold is increased over 65\%, it reduces the recall rate which affects accuracy and F-score. The evaluation indicates that the threshold value of 60\% provides an accuracy of 97\% and F-score of 80\%. Also, From Figure~\ref{fig}, one can observe the relation between FPR and TPR of Naive Bayes-based detection. For FPR larger than 0.33, TPR becomes 1.

For Naive Bayes-based detection on the smartphone, we considered nine activities in total (three general activities and six user-related activities)~\cite{203854}. From Table~\ref{naivetable}, we can observe that TP rate FP rates decrease with the threshold value while FN and TN increase. When the threshold is 60\%, the peak accuracy (95\%) and F-score (82\%) are achieved for the smartphone. Precision-Recall curve for Naive Bayes model is given in Figure~\ref{fig:f4}. We can notice that PRC curve is more irregular compared to Markov Chain-based approach. Calculated auPRC for Naive Bayes-based approach is 0.7 for the smart watch and the smartphone, both of which indicate less efficient method for imbalanced dataset.

\vspace{-0.4cm}
\subsection{Evaluation of Alternative Detection Techniques}
In alternative detection techniques, we used other supervised machine learning techniques to train the 6thSense framework for both the smart watch and the smart phone. For this, we utilized WEKA and it provides three types of analysis - split percentage analysis, cross-validation analysis, and supplied test set analysis. We chose 10 fold cross-validation analysis to ensure that all the data was used for both training and test. Thus, the error rate of the predictive model would be minimized in the cross validation. In Table~\ref{wekatable}, a detailed evaluation of different machine learning algorithms is given for 6thSense. For \textit{Rule Based Learning}, 6thSense has the best result for \textit{PART} algorithm, which has an accuracy of 0.98 and F-score of 0.80. On the other hand, for \textit{Regression Analysis}, we use the logistic function which has high FPR (0.65) and lower F-score (0.49). 
\textit{Multilayer Perceptron} algorithm gives an accuracy of 0.9878 and F-score of 0.80,  which is higher than previously mentioned algorithms. However, FPR is much higher (0.35), which is actually a limitation for intrusion detection frameworks in general. Compared to these algorithms, \textit{Linear Model Tree (LMT)} gives better results in detecting sensor-based attacks. This evaluation indicates that \textit{LMT} provides an accuracy of 0.99 and F-score of 0.972 for the smart watch.\par
From Table~\ref{wekatable}, one can also see performance of different machine learning algorithms in 6thSense on a smartphone. Here, LMT achieves the highest accuracy and F-score of 0.99 and 0.96, respectively. Multilayer Perception algorithm also performs well with F-score of 0.82. However, false positive rate is high in this algorithm which decreases the performance. In summary, LMT works efficiently in both the smart watch and the smart phone.
\vspace{-0.5cm}
\subsection{Comparison of Detection Methods}
In this subsection, we give a comparison among the different machine learning-based detection approaches tested in 6thSense for defending against sensor-based threats on the smart watch and the smartphone. For all the approaches, we select the best possible case and report their performance metrics.\par Table~\ref{comparison} depicts comparison among different detection approaches on the smart watch. For Markov Chain-based detection, we chose three consecutive malicious states as valid device conditions. On the other hand, in Naive Bayes approach, the best performance is observed for the threshold of 60\%. For other machine learning algorithms tested via WEKA, we chose LMT as it gives highest accuracy among other machine learning algorithms. These results indicate that both Markov Chain and LMT provide high accuracy and F-score compared to the Naive Bayes-based approach.
\par On the contrary, Naive Bayes model displays higher recall rate and less FNR than other approaches. However, the presence of FPR in IDS is an issue to the system since FPR refers to a malicious attack that is identified as a valid device state. Both Markov Chain and LMT has lower FPR. Again, as our dataset is imbalanced (number of benign activities is higher than malicious activity), we chose auPRC as one of the performance metric of 6thSense. From Table~\ref{comparison} we can see that Markov Chain-based detection has the highest auPRC (0.926) followed by LMT (0.892) and Naive Bayes (0.646). In summary, considering F-score, accuracy, and auPRC of all three approaches, we conclude that Markov Chain and LMT both performs effectively for 6thSense.

In Table~\ref{comparison}, we present a comparison of different machine learning-based detection techniques used in 6thSense on the smartphone. Again, we chose the best possible (Markov Chain, Naive Bayes, and LMT) cases for all of the approaches and compare them in Table~\ref{comparison}. Similar to
results in the smart watch, threshold for Markov Chain-based detection is three consecutive malicious state. For Naive Bayes-based detection, best performance can be observed for 60\% threshold probability. From Table~\ref{comparison}, we can observe that Markov Chain and LMT performs with high accuracy and F-score compared to Naive Bayes-based approach. Naive Bayes model also introduces high FP rate (0.3) which indicates poor performance for IDS. On the contrary, Markov Chain and LMT shows lower FP rate (0 and 0.0694 respectively). Again, from Figure~\ref{fig:f4}, we can observe that Naive Bayes model has low auPRC compared to Markov Chain-based detection in Figure~\ref{fig:f2}. LMT also has high auPRC (0.91) which is suitable for our imbalanced dataset. In summary, both Markov Chain and LMT performs well for 6thSense on the smart phone with high accuracy, F-score, and auPRC.
\begin{table}[h!]
\centering
\fontsize{8}{10}\selectfont
\resizebox{0.4\textwidth}{!}{
\begin{tabular}{lllll}
\toprule
\textbf{\begin{tabular}[c]{@{}l@{}}Performance \\Metrics\end{tabular}} & \textbf{\begin{tabular}[c]{@{}l@{}}Markov Chain\\ (Smart Watch$\vert$\\Smart Phone) \end{tabular}} & \textbf{\begin{tabular}[c]{@{}l@{}}Naive Bayes\\(Smart Watch$\vert$\\Smart Phone)\end{tabular}}  & \textbf{\begin{tabular}[c]{@{}l@{}}LMT\\(Smart Watch$\vert$\\Smart Phone)\end{tabular}}\\ \hline
\midrule
Recall rate         & 0.9770  $\vert$  0.98    &  1   $\vert$   1    & 0.9998   $\vert$   0..998             \\ 
\begin{tabular}[c]{@{}l@{}}False Negative\\ Rate \end{tabular}& 0.0230  $\vert$   0.02   & 0  $\vert$  0       & 0.0002   $\vert$   0.0002            \\
Precision rate      & 0.9857  $\vert$   1      & 0.67  $\vert$   0.7    & 0.9458    $\vert$   0.9306           \\ 
\begin{tabular}[c]{@{}l@{}}False positive \\rate \end{tabular}& 0.0143   $\vert$  0      & 0.33   $\vert$  0.3    & 0.0694  $\vert$     0.07           \\
Accuracy            & 0.9795  $\vert$  0.9833  & 0.9720  $\vert$  0.9492   & 0.998  $\vert$  0.9997             \\
F-Score             & 0.9813  $\vert$  0.9899  & 0.80  $\vert$  0.8235  & 0.972   $\vert$  0.964             \\
auPRC               & 0.926   $\vert$  0.947  & 0.646  $\vert$  0.686   & 0.892  $\vert$   0.91               \\
\bottomrule
\end{tabular}}
\caption{Comparison of different machine-learning-based approaches proposed for 6thSense on Smartwatch and Smartphone (i.e., Markov Chain, Naive Bayes, and LMT).}
\label{comparison}
\end{table}
\vspace{-0.6cm}
\subsection{Performance Overhead}
As previously mentioned, 6thSense collects data in an Android device from different sensors (permission and no-permission imposed sensors). In this sub-section, we measure the performance overhead introduced by 6thSense on the tested Android devices (smart watch and smartphone) in terms of CPU usage, RAM usage, file size, and power consumption. Table~\ref{overhead}, Table~\ref{analysis}, and Table~\ref{analysis_phone} give the details of the performance overhead of 6thSense on the smart watch and the smartphone.

For no permission-imposed sensors, the data collection phase logs all the values within a time interval which causes an increased usage of RAM, CPU and Disc compared to permission-imposed sensors. For the power consumption, we observe that no permission-imposed sensors use higher power than permission-imposed sensors. This is mainly because permission-imposed sensors are logic-oriented and have lower sampling rate, which reduces its resource needs.
\begin{table}[h!]
\vspace{-0.2cm}
\centering
\fontsize{8}{10}\selectfont
\resizebox{0.4\textwidth}{!}{
\begin{tabular}{p{2cm}p{1cm}p{2.2cm}p{2.6cm}}
\toprule
\textbf{Parameters}                                                                    & \textbf{Time}       & \textbf{\begin{tabular}[c]{@{}l@{}}No-permission \\ imposed sensors\\(Smart Watch$\vert$\\Smart Phone)\end{tabular}} & \textbf{\begin{tabular}[c]{@{}l@{}}Permission-imposed\\sensors (Smart Watch$\vert$\\Smart Phone)\end{tabular}} \\ \hline
\midrule
CPU Usage   & N/A   & 5.5\%  $\vert$  3.9\%     & 2.5\%  $\vert$  0.3\%    \\ \hline
{\begin{tabular}[c]{@{}l@{}}RAM Usage\\(MB)\end{tabular}} & N/A   & 17  $\vert$  23      & 11  $\vert$  14      \\ \hline
\multirow{3}{*}{\begin{tabular}[c]{@{}l@{}}Disc Usage\\(MB)\end{tabular}}     & 1 min  & 4  $\vert$  6.5   & 0.001  $\vert$  0.001   \\ \cline{2-4} 
                                & 5 min  & 7.5  $\vert$  9   & 0.001  $\vert$  0.002   \\ \cline{2-4} 
                                & 10 min & 10  $\vert$  12   & 0.001  $\vert$  0.003   \\ \hline
\multirow{3}{*}{\begin{tabular}[p{1.7cm}]{@{}p{2cm}@{}}Power\\Consumption (mW)\end{tabular}} 
                                & 1 min  & 10.5  $\vert$  13.5  & 2  $\vert$  3.12 \\ \cline{2-4} 
                                & 5 min  & 45.6  $\vert$  96.67 & 16.5  $\vert$  27.4 \\ \cline{2-4} 
                                & 10 min & 78.4  $\vert$  133.33 & 27  $\vert$  45     \\ \hline
\multirow{3}{*}{\begin{tabular}[p{1.7cm}]{@{}p{2cm}@{}}Power\\Consumption\\(without data file)\end{tabular}} 
                                & 1 min  & 1.32  $\vert$  2.68  & 0.1  $\vert$  0.23  \\ \cline{2-4} 
                                & 5 min  &  8.7  $\vert$  23.4  & 2 $\vert$ 9.63 \\ \cline{2-4} 
                                & 10 min & 32.56 $\vert$  55.35 & 9 $\vert$ 17 \\ \\\hline
\bottomrule                                                                              
\end{tabular}}
\vspace{-0.1cm}
\caption{Performance overhead of data collection.}
\vspace{-0.2cm}
\label{overhead}
\end{table}
The overall performance overhead is as low as 5.5\% of CPU, less then 17MB RAM space, and less than 10MB disc space for the smart watch. Compared to the smart watch, performance overhead for the smartphone is higher because of higher number of sensors. Nevertheless, smartphone offers more resources (CPU speed, RAM size, disc size) than the smart watch which minimizes the effect of performance overhead. Performance overhead for the smartphone is as low as 3.9\%, less than 6.5MB RAM space, and less than 12MB disc space. Thus, 6thSense's overhead is minimal and acceptable for an IDS system on current smart devices. One of the main concerns of implementing 6thSense on smart device is the power consumption.

Table~\ref{overhead} also shows the power consumption of the Android app used in 6thsense. For one minute, 6thsense consumes 10.5 mW power which increases upto 78.4 mW for ten minutes on a smartwatch. For a smartphone, 6thSense consumes upto 133.33 mW power for ten minutes. The main reason of this high power consumption is that all the sensors are kept on for the data collection and all the data are saved on device for later analysis. To mitigate power-performance trade-off, in practical settings, the data are not saved on device rather a real-time analysis is done, which indeed decreases the power consumption. Without saving the data, the power consumption significantly becomes smaller. From Table~\ref{overhead}, we can observe that the power consumption of 6thSense becomes 32.56 mW which is almost 2 times lower than otherwise on a smartwatch. For real-time analysis, power consumption of 6thSense decreases 2.4 times on the smartphone. As all the sensors do not have to remain on for the analysis part, data can be observed if the smart device is in unlocked status to lower the power consumption.  

\par Moreover, for the data analysis phase of 6thSense, we also implemented Markov Chain, Naive Bayes, and LMT-based detection methods on the Android smartphone and smart watch.  Table~\ref{analysis} shows the performance overhead of different detection techniques used in 6thSense on a smart watch. All three detection techniques yield less than 2\% CPU usage and 10 MB of RAM usage. Note that we consider the disc usage as a performance overhead for the data analysis phase since results can be stored for further performance evaluation of the framework. Our extensive evaluation shows that the disc usage for the data analysis of 6thSense is less then 1 MB in all the three detection methods for 5 minutes of analysis. Table~\ref{analysis} also provides the power consumption of different detection techniques of 6thSense. We can observe that the power consumption of the data analysis phase is comparatively lower (less than 5 mW) than the data collection phase of 6thSense. Finally, Table~\ref{analysis_phone} shows performance evaluation of different detection techniques of 6thSense on an Android operated smartphone. 6thSense performs with minimum overhead with less than 3\% CPU usage, 17 MB RAM usage, and 2 MB of disc usage. Power consumption in the smartphone is also as low as 6 mW for different detection techniques implemented on 6thSense. In summary, different detection methods used in 6thSense offer lower performance overhead in the system.
\begin{table}[t!]
\centering
\vspace{6pt}
\fontsize{8}{10}\selectfont
\resizebox{0.35\textwidth}{!}{
\begin{tabular}{p{2cm}p{1.6cm}p{1.6cm}p{1.5cm}}
\toprule
\textbf{Parameters}       & \textbf{\begin{tabular}[c]{@{}l@{}}Markov Chain\end{tabular}} & \textbf{\begin{tabular}[c]{@{}l@{}}Naive Bayes\end{tabular}} & \textbf{\begin{tabular}[c]{@{}l@{}}LMT\end{tabular}}\\ \hline
\midrule
CPU Usage       & 1\%        & 1.5\%    & 1\%                                                                 \\ \hline
RAM Usage       & 6 MB        & 10 MB      & 10 MB                                                                 \\ \hline
\begin{tabular}[c]{@{}l@{}}Disc Usage\\(For 5 Min)\end{tabular}      & \textless 1 MB       & \textless 400 kB    & \textless 400 kB   \\\hline
\begin{tabular}[p{1.7cm}]{@{}p{2cm}@{}}Power\\Consumption\\(For 5 min)\end{tabular} & 1 mW      & 2 mW & 2.5 mW      \\ \hline
\bottomrule                                                                              
\end{tabular}}
\caption{Performance overhead of the data analysis phase in 6thSense on smart watch.}
\vspace{-10pt}
\label{analysis}
\end{table}
\begin{table}[h!]
\centering
\fontsize{8}{10}\selectfont
\resizebox{0.35\textwidth}{!}{
\begin{tabular}{p{2cm}p{1.6cm}p{1.6cm}p{1.5cm}}
\toprule
\textbf{Parameters}       & \textbf{\begin{tabular}[c]{@{}l@{}}Markov Chain\end{tabular}} & \textbf{\begin{tabular}[c]{@{}l@{}}Naive Bayes\end{tabular}} & \textbf{\begin{tabular}[c]{@{}l@{}}LMT\end{tabular}}\\ \hline
\midrule
CPU Usage       & 1.2\%        & 2.5\%    & 1\%                                                                 \\ \hline
RAM Usage       & 12 MB        & 15 MB      & 17 MB                                                                 \\ \hline
\begin{tabular}[c]{@{}l@{}}Disc Usage\\(For 5 Min)\end{tabular}      & \textless 2 MB       & \textless 1 MB    & \textless 1 MB   \\\hline
\begin{tabular}[p{1.7cm}]{@{}p{2cm}@{}}Power\\Consumption\\(For 5 min)\end{tabular} & 4.5 mW      & 6 mW & 3 mW      \\ \hline
\bottomrule                                                                              
\end{tabular}}
\caption{Performance overhead of the data analysis phase in 6thSense on smartphone.}
\vspace{-0.7cm}
\label{analysis_phone}
\end{table}

\subsection{Power-efficiency trade-off} 
One major concern of implementing a security framework in smart devices is power-efficiency trade-off. As smart devices such as smart watch and smart phone are resource constrained devices, an efficient security framework should work accurately with limited resources. 6thSense uses all the available sensors in a device to understand the state of the device and detects sensor-based threat based on state transition model. This can be a drawback in terms of power consumption of the device. To address this limitation, we performed a power-frequency trade-off study to determine the working condition of 6thSense in real-life settings. According to Nielsen, average American adult spends around 3 hours everyday on their smartphone~\cite{nielsen}.

\begin{figure}[h!]
\vspace{-20pt}
    \centering
    \subfloat[]{\includegraphics[width=0.25\textwidth]{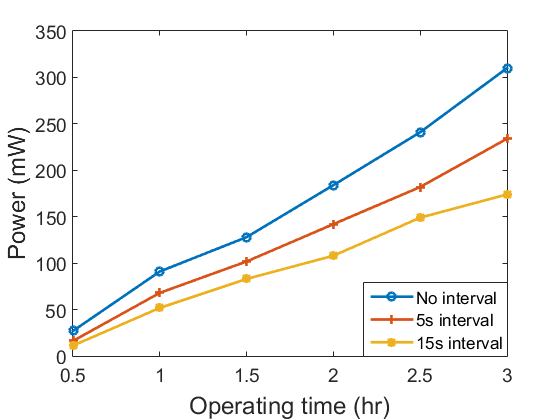}\label{fig:f5}}
    \subfloat[]{\includegraphics[width=0.25\textwidth]{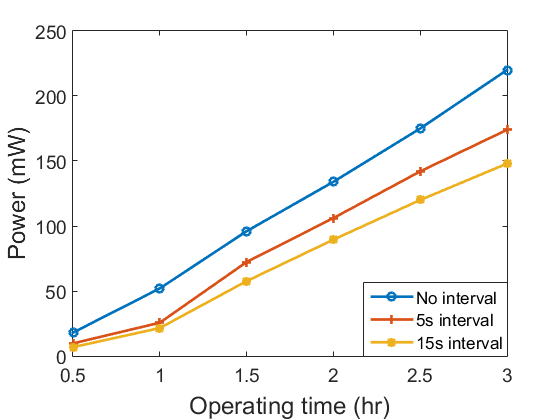}\label{fig:f6}}
    \vspace{-0.2cm}
    \caption{Power-frequency for (a) smart watch (b) smartphone}
    \label{power-freq}
    \vspace{-0.2cm}
\end{figure}

We consider this as an average time that 6thSense has to run to detect any sensor-based threats in a smart device. In Figure~\ref{fig:f5} and~\ref{fig:f6}, we illustrate the average power consumption graph for 6thSense with different scanning frequency in a smartphone and a smart watch, respectively. One can notice that 6thSense consumes 310 mW power for scanning continuously for 3 hours in a smart phone (Figure~\ref{fig:f5}). Average power consumption lowers to 234 and 174 mW with 5s and 15s time interval respectively. For smart watch, highest average power consumption for 6thSense is 220 mW for continuous scan. Average power consumption becomes as low as 174 mW and 148 mW for 5s and 15s time interval respectively.
\vspace{-10pt}
\subsection{Discussion and Limitations}
\begin{itemize}
    \item \textit{Sensor-based threats in real-life settings:} One limitation of 6thSense is the adversaries (sensor-based attacks) used in the evaluation were constructed in a lab-environment. Note that as of this writing there are no real sensor-based malware in the wild. However, many independent researchers have confirmed the feasibility of sensor-based threats for smart devices~\cite{sikder2018survey}. Indeed, more recently, ICS-CERT also warned the vendors and the wider communities about the possibility of exploiting the sensors of a device to alter sensors’ output in a controlled way to perform malicious behavior in the device~\cite{ICS-CERT}. Google also acknowledges the sensor-based threats by restricting sensor access for background apps in version of Android~\cite{andrp}. To understand the sensor-based threats and limitation of existing solutions, we built the proof-of-concept versions of the sensor-based threats discussed in Section 4. We also note that to ensure the reliability of the malware (i.e., specific malicious Apps) for the threats described in Section 4, we checked how they perform with respect to the real malicious software scanners. For this, we uploaded our malware on \textit{VirusTotal} and tabulated the results of the performance of 60 different malware scanners available at the VirusTotal website in Table~\ref{scan}. As seen in this table, the sensor-based threats are not recognized by the different scanners. \textit{In conclusion, current malware scanners are not aware of these threats yet and our malware can be reliably used to test the efficiency of 6thSense.}
\begin{table}[h!]
\vspace{-0.1cm}
\centering
\fontsize{8}{10}\selectfont
\resizebox{0.25\textwidth}{!}{
\begin{tabular}{cc}
\toprule
\textbf{Adversary Model} & \textbf{Detection Ratio} \\ \hline
\midrule
Threat-1 & 0/60 \\\hline
Threat-2 & 0/60 \\\hline
Threat-3 & 0/62 \\\hline
\bottomrule                                     
\end{tabular}}
\vspace{-3pt}
\caption{\textit{VirusTotal} scan result for the adversary models.}
\label{scan}
\vspace{-0.2cm}
\end{table}
    \item \textit{Power monitoring app:} Different smartphone and smart watch vendors offer power monitoring apps which monitor running apps (both background and foreground apps) and minimize the power consumption of the device. For example, Samsung provides a power monitoring app that prevents background apps to drain power. Power monitoring apps activate sleep mode which disables the updates and notifications for the inactive apps. This conflicts with the malicious apps described in Section 7.1. However, the power monitoring app only works when the app is in the background. If a foreground app has malicious sensor logic, it can easily bypass the power monitoring app and initiate an attack. As power monitoring apps restrict important updates (e.g., messages from text apps, alarm apps, etc.), users can turn-off or modify this feature for convenience~\cite{powermon}. Moreover, the smart watches do not have any power monitoring option which makes them vulnerable to sensor-based threats. \textit{In summary, power monitoring app can restrict sensor-based threats to some extent, but can not nullify them entirely.}  
    \item \textit{New OS feature:} Recently, Android introduced a new version of OS (Android P) which restricts camera and microphone usage if an app runs in the background. This feature certainly acknowledges the sensor-based threats and restricts sensor misuse in a smartphone. However, Android P only eliminates one threat model described in Section 4 and 7.1. Different malicious apps can still access other sensors in the background and perform multiple malicious activities. As explained in Section 7.1, Threat Model-1 uses motion and light sensors which does not have any conflict with Android P. Threat Model-2 uses the microphone of a connected smart watch which bypasses the security feature of Android P. Threat Model-3 triggers the camera of a smartphone if all the other sensors are inactive. Here, the malicious app opens the camera in the foreground which is allowed by Android P. We developed an updated version of this attack which could start recording via microphone in a smart watch if the connected smartphone was inactive and thus, bypass the security feature of Android P. Again, Android P only nullifies the threat if the app is installed in a smart phone. A malicious app installed in a smart watch can trigger the camera of a smart phone without any restriction. Also, only 1\% of Android-powered devices support Android P currently which makes majority of the devices vulnerable to sensor-based threats using camera or microphone surreptitiously~\cite{andrp2}. \textit{In short, even with the introduction of the new OS, sensor-based threats can still affect normal operations of the smart devices.}
    \item \textit{Optimum scanning frequency:} As smart devices are resource-constrained devices, an optimum scanning frequency is needed for 6thSense to lower the power consumption of the device. In Section 7.8, we illustrated that by scanning the sensors in fixed intervals (15s) and unlocked states, power consumption can be lowered by approximately 43\%. However, some sensor-based threats can bypass the lock state and perform malicious activities in smart devices. To address this limitation, 6thSense can use the context-aware model to detect the lock state of the device and monitor limited sensors to minimize the power consumption. As Android P is restricting some sensors (microphone and camera), 6thSense can use this feature and select limited sensors to scan in the locked state. \textit{In short, performance of 6thSense can be configured in terms of power consumption by selecting optimum scanning frequency and combining with existing permission model of OS.} 
\end{itemize}
\vspace{-0.2cm}
\par

\vspace{-0.2cm}
\section{Conclusion}\label{sec:Conclusion}
Wide utilization of sensor-rich smart devices created a new attack surface namely sensor-based attacks. Accelerometer, gyroscope, light, etc. sensors can be abused to steal and leak sensitive information or malicious Apps can be triggered via sensors. Security in current smart devices lacks appropriate defense mechanisms for such sensor-based threats. In this paper, we presented 6thSense, a novel context-aware task-oriented sensor-based attack detector for smart devices. We articulated problems in existing sensor management systems and different sensor-based threats for smart devices. Then, we presented the design of 6thSense to detect sensor-based attacks on sensor-rich smart devices (smartwatch and smartphone) with low-performance overhead. 6thSense utilized different machine learning (ML) techniques to distinguish malicious activities from benign activities on a device. To the best of our knowledge, 6thSense is the first comprehensive context-aware security solution against sensor-based threats. We evaluated 6thSense on real devices with 100 different individuals. 6thSense achieved over 97\% of accuracy with different ML algorithms including Markov Chain, Naive Bayes, and LMT. We also evaluated 6thSense against three different sensor-based threats, i.e., information leakage, eavesdropping, and triggering a malware via sensors. The empirical evaluation revealed that 6thSense is highly effective and efficient at detecting sensor-based attacks while yielding minimal overhead. 
\vspace{4pt}

\vspace{-0.5cm}
\section*{Acknowledgments}

This work was partially supported by US NSF-CAREER-CNS-1453647, NSF-CNS-1718116, and Florida Center for Cybersecurity Capacity Building Program.  The  views  expressed are  those  of  the authors only.

\vspace{-0.2cm}
\ifCLASSOPTIONcaptionsoff
  \newpage
\fi

\bibliographystyle{IEEEtran}
\vspace{-0.3cm}
\bibliography{sample}
\appendices
\vspace{-1cm}
\begin{IEEEbiography}[{\includegraphics[width=1in,height=2in,clip,keepaspectratio]{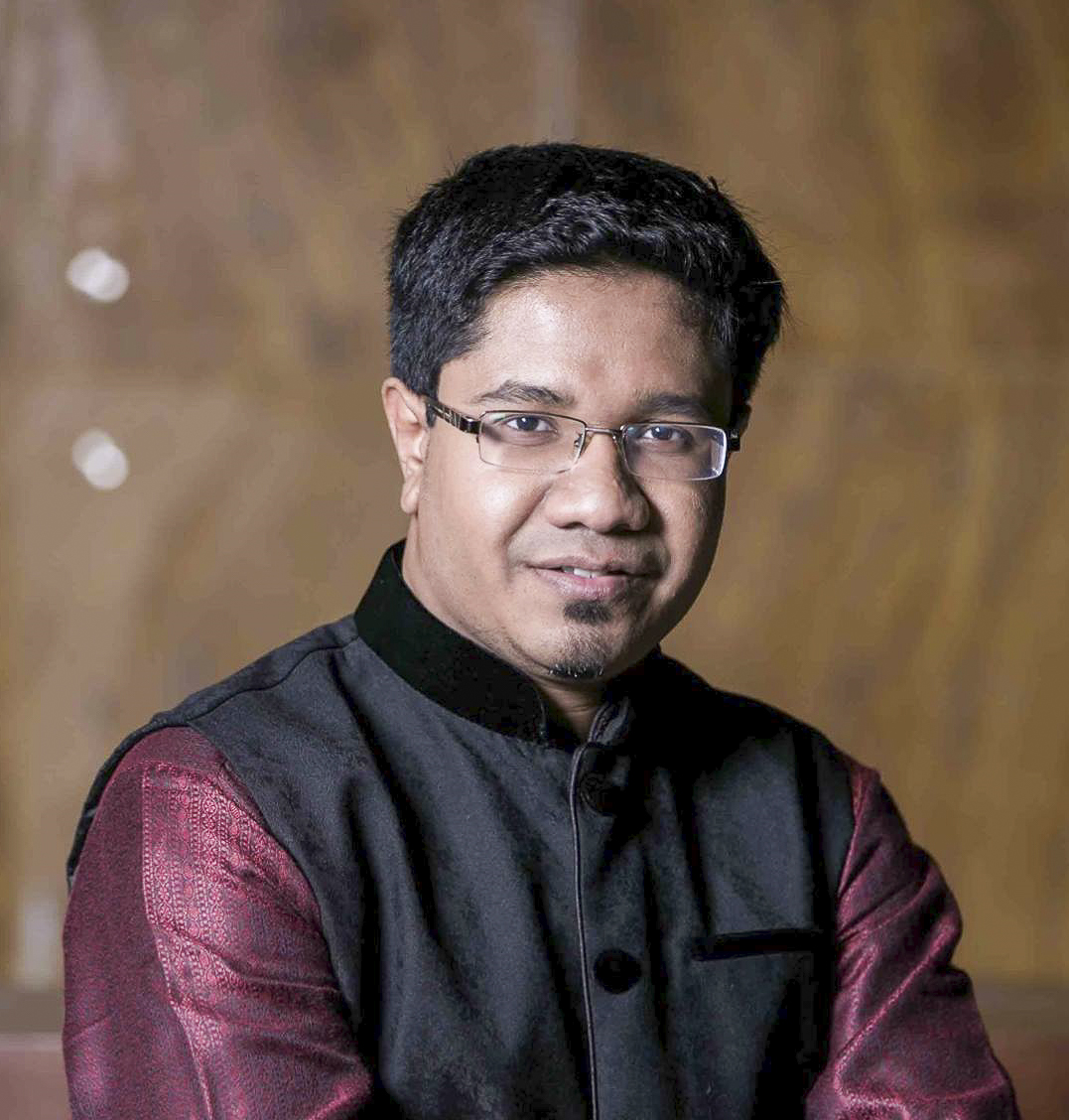}}]{Amit Kumar Sikder}
 is currently a PhD student and Research Assistant in the Department of Electrical and Computer Engineering at Florida International University, as a member of the Cyber-Physical Systems Security Lab (CSL). He previously completed his Bachelors in Electrical and Electronic Engineering from Bangladesh University of Engineering and Technology (BUET). His research interests are focused on the security of Cyber-Physical Systems (CPS) and Internet of Things (IoT). He also has worked in areas related to security of smart devices, security of smart home, smart city, wireless communication. More information can be obtained from: http://web.eng.fiu.edu/asikd003/. 
\end{IEEEbiography}
\vspace{-1cm}
\begin{IEEEbiography}[{\includegraphics[width=1in,height=2in,clip,keepaspectratio]{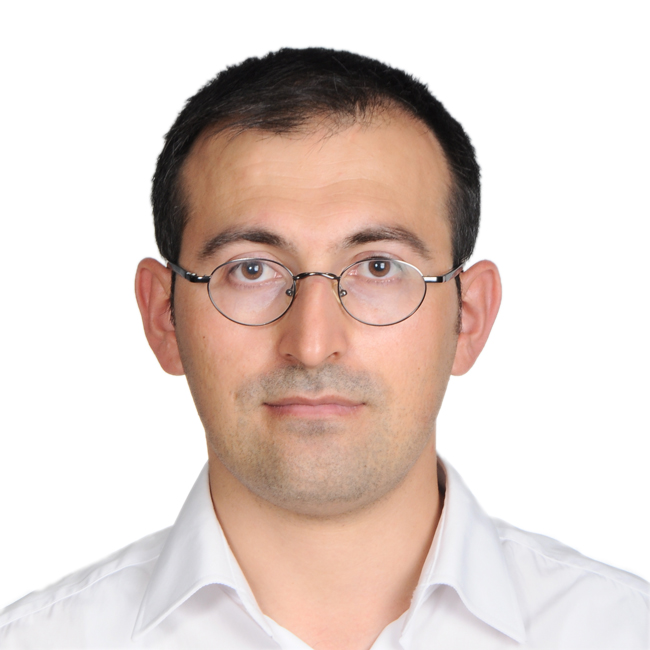}}]{Hidayet Aksu}
received his Ph.D., M.S. and B.S. degrees from Bilkent University, all in Department of Computer Engineering, in 2014, 2008 and 2005, respectively. He is currently a Postdoctoral Associate in the Department of Electrical \& Computer Engineering at Florida International University (FIU). Before that, he worked as an Adjunct Faculty in the Computer Engineering Department of Bilkent University. He conducted research as visiting scholar at IBM T.J. Watson Research Center, USA in 2012-2013. He also worked for Scientific and Technological Research Council of Turkey (TUBITAK). His research interests include security for cyber-physical systems, internet of things, security for critical infrastructure networks, IoT security, security analytics, social networks, big data analytics, distributed computing, wireless networks, wireless ad hoc and sensor networks, localization, and p2p networks.
\end{IEEEbiography}
\vspace{-1cm}
\begin{IEEEbiography}[{\includegraphics[width=1in,height=1.2in,clip]{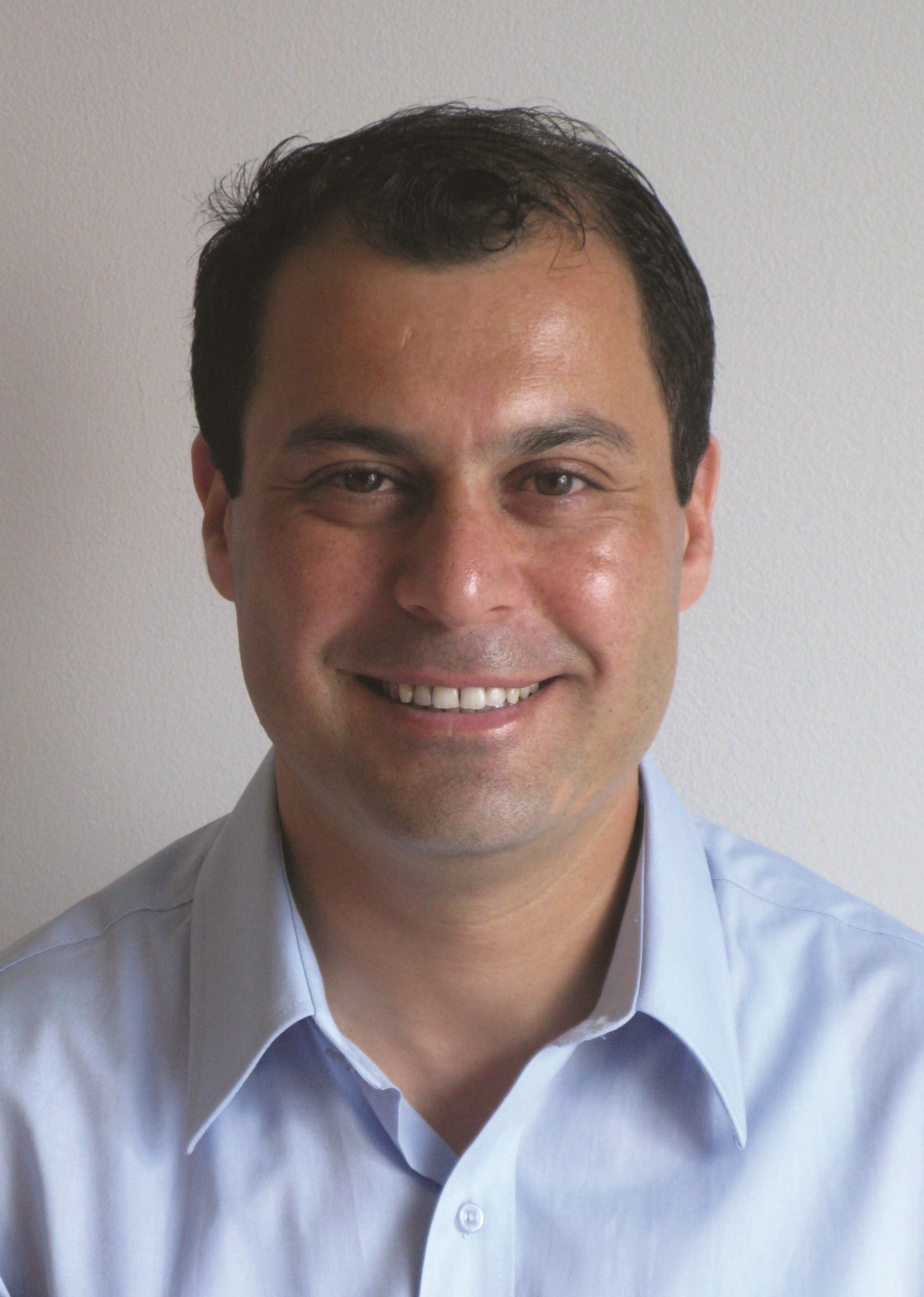}}]{Dr. A. Selcuk Uluagac} is currently an Assistant Professor in the Department of Electrical and Computer Engineering (ECE) at Florida International University (FIU). Before joining FIU, he was a Senior Research Engineer in the School of Electrical and Computer Engineering (ECE) at Georgia Institute of Technology. Prior to Georgia Tech, he was a Senior Research Engineer at Symantec. He earned his Ph.D. with a concentration in information security and networking from the School of ECE, Georgia Tech in 2010. He also received an M.Sc. in Information Security from the School of Computer Science, Georgia Tech and an M.Sc. in ECE from Carnegie Mellon University in 2009 and 2002, respectively. 
The focus of his research is on cyber security topics with an emphasis on its practical and applied aspects. He is interested in and currently working on problems pertinent to the security of Cyber-Physical Systems and Internet of Things. In 2015, he received a Faculty Early Career Development (CAREER) Award from the US National Science Foundation (NSF). In 2015, he was awarded the US Air Force Office of Sponsored Research (AFOSR)'s 2015 Summer Faculty Fellowship. In 2016, he received the Summer Faculty Fellowship from the University of Padova, Italy. He is also an active member of IEEE (senior grade), ACM, and ASEE and a regular contributor to national panels and leading journals and conferences in the field. Currently, he is the area editor of Elsevier Journal of Network and Computer Applications and serves on the editorial board of the IEEE Communication Surveys and Tutorials. More information can be obtained from: http://web.eng.fiu.edu/selcuk.
\end{IEEEbiography}
\end{document}